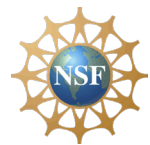

# QUANTUM INFORMATION AND COMPUTATION FOR CHEMISTRY

## 2016

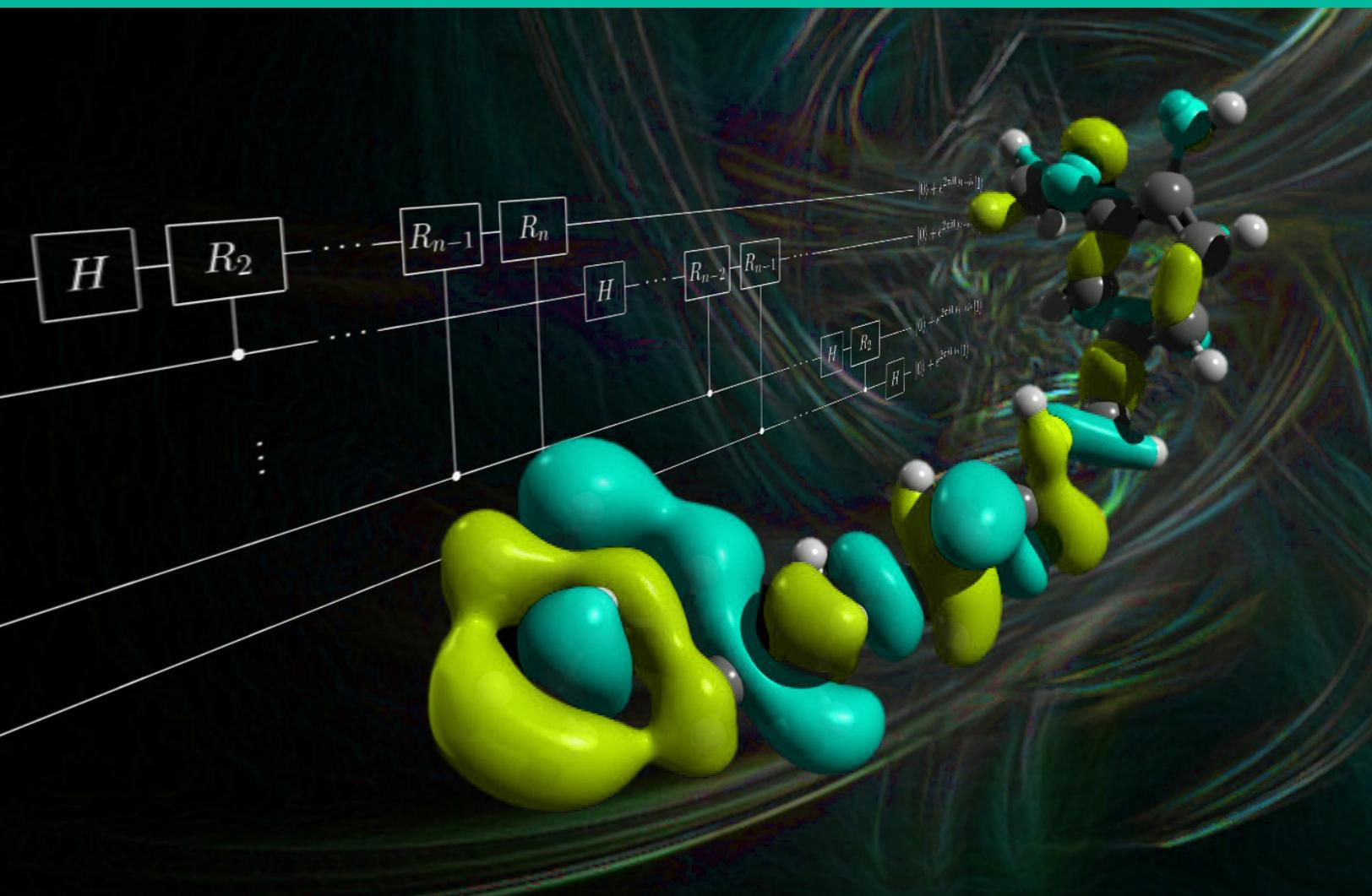

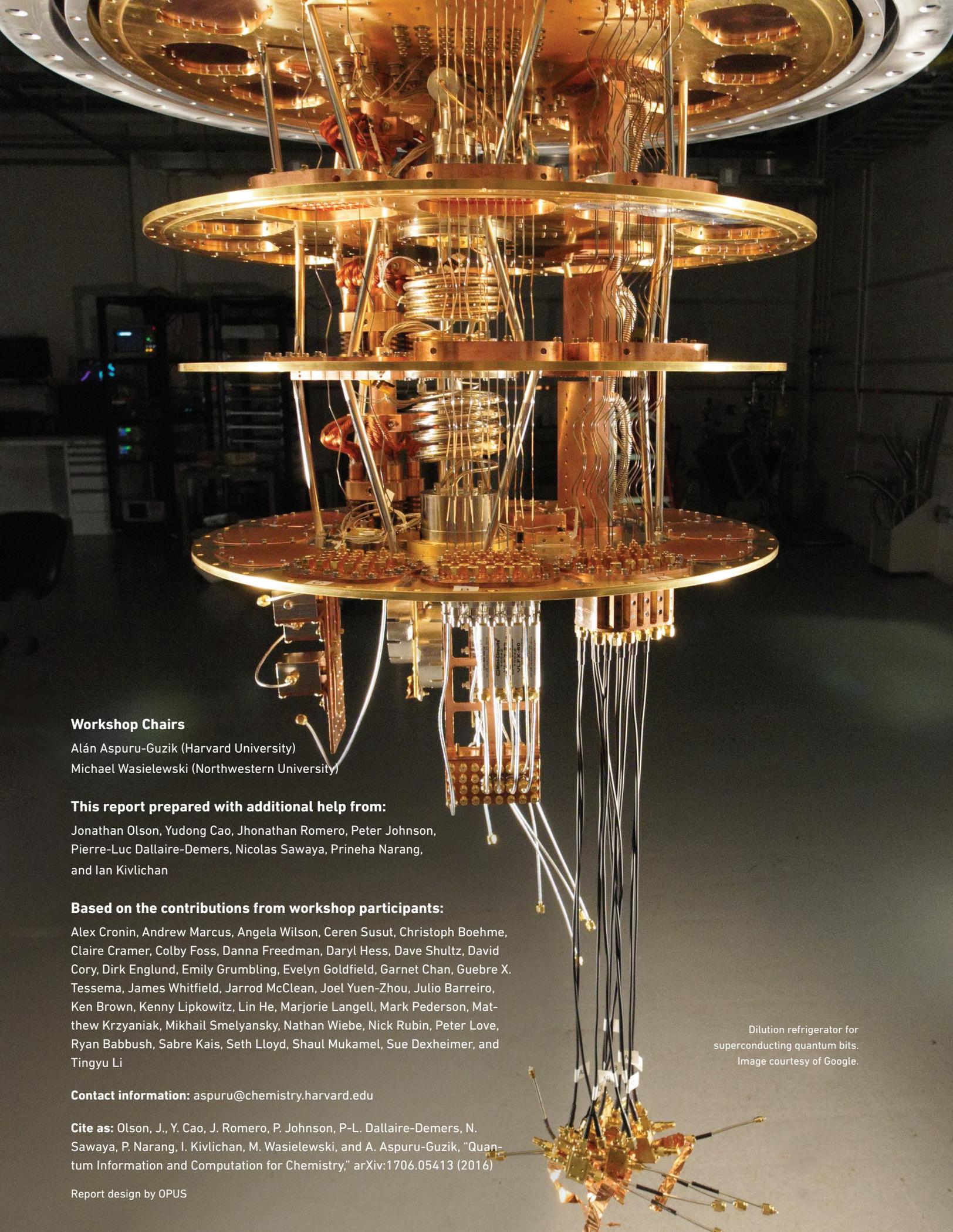

**Workshop Chairs**

Alán Aspuru-Guzik (Harvard University)
Michael Wasielewski (Northwestern University)

**This report prepared with additional help from:**

Jonathan Olson, Yudong Cao, Jhonathan Romero, Peter Johnson,
Pierre-Luc Dallaire-Demers, Nicolas Sawaya, Prineha Narang,
and Ian Kivlichan

**Based on the contributions from workshop participants:**

Alex Cronin, Andrew Marcus, Angela Wilson, Ceren Susut, Christoph Boehme,
Claire Cramer, Colby Foss, Danna Freedman, Daryl Hess, Dave Shultz, David
Cory, Dirk Englund, Emily Grumbling, Evelyn Goldfield, Garnet Chan, Guebre X.
Tessema, James Whitfield, Jarrod McClean, Joel Yuen-Zhou, Julio Barreiro,
Ken Brown, Kenny Lipkowitz, Lin He, Marjorie Langell, Mark Pederson, Mat-
thew Krzyaniak, Mikhail Smelyansky, Nathan Wiebe, Nick Rubin, Peter Love,
Ryan Babbush, Sabre Kais, Seth Lloyd, Shaul Mukamel, Sue Dexheimer, and
Tingyu Li

**Contact information:** aspuru@chemistry.harvard.edu

**Cite as:** Olson, J., Y. Cao, J. Romero, P. Johnson, P-L. Dallaire-Demers, N.
Sawaya, P. Narang, I. Kivlichan, M. Wasielewski, and A. Aspuru-Guzik, "Quan-
tum Information and Computation for Chemistry," arXiv:1706.05413 (2016)

Report design by OPUS

Dilution refrigerator for
superconducting quantum bits.
Image courtesy of Google.

# TABLE OF CONTENTS

# 1 INTRODUCTION

## 1.1 BACKGROUND

*...Hope for simulating increasingly complex quantum systems lies in a new paradigm for information processing: quantum computation.*

Modern computational chemistry drives innovation in areas ranging from drug discovery to material design. These application-fueled fields demand rapid development of analytic and computational tools. While technological advances have continued to follow the steady growth of computing power, the potential for a disruptive technology is emerging at the interface of two fields: computational chemistry and quantum information science.

In order to accurately model chemical systems, we must account for their inherent quantum nature. Despite significant advances in the field of molecular simulation, many important problems in chemistry such as the prediction of chemical reactions and the description of excited electronic states, transition states, and ground states of transition metal complexes, which are necessary for nearly all aspects of material design, remain challenging. Approximation methods motivated by classical physics allow for efficient simulation of certain quantum systems, but fail to capture the physics of such strongly correlated quantum systems. In many cases, simulating these strongly correlated quantum systems to a desirable accuracy using current methods requires computational resources which scale exponentially with system size.

As envisioned by the great 20[th] century physicist Richard Feynman, hope for simulating increasingly complex quantum systems lies in a new paradigm for information processing: quantum computation. Such computers can store and process information about simulated quantum systems natively, reducing the computational resource scaling with the size of the system to just polynomial growth, in principle. In the thirty years since Feynman's forecast, the field of quantum information science has emerged and has made tremendous strides: experimental advances in controlling quantum systems have brought quantum computers to the brink of outperforming classical computation; the synergy between quantum chemistry and quantum algorithm development has continued to equip and refine the toolbox of disruptive applications for quantum computation; and at the interface between computational chemistry and quantum information science, researchers are poised to carry Feynman's vision to fruition.

## 1.2 OBJECTIVE

The NSF Workshop in Quantum Information and Computation for Chemistry assembled experts from directly quantum-oriented fields such as algorithms, chemistry, machine learning, optics, simulation, and metrology, as well as experts in related fields such as condensed matter physics, biochemistry, physical chemistry, inorganic and organic chemistry, and spectroscopy.

The goal of the workshop was to identify target areas where cross fertilization among these fields would result in the largest payoff for developments in theory, algorithms, and experimental techniques. The outcome of the workshop is underscored in the agenda summarized below:

- Develop specific quantum algorithms that are relevant for chemistry.

- Identify the potential impact of quantum machine learning, quantum sensing, and quantum communication on chemistry.

- Create novel platforms for implementing specific quantum algorithms and simulation.

- Translate the tools of quantum information into the language of chemistry.

- Identify the experimental challenges for which chemistry can enable the next generation of quantum devices.

- Bring the scientific communities of chemistry and quantum information together.

A quantum information perspective can provide new tools for theoretical and computational chemistry. From the theoretical aspect, the understanding of chemical phenomena within the framework of quantum information may enlighten our understanding of chemistry by approaching these problems with a different toolset. From the computational perspective, the potential for solving Schrödinger's equation in a *numerically exact* fashion will allow for a transformation of our ability to simulate chemical systems. This in turn, will lead to an enhanced ability to explore chemical space.

A quantum information perspective can provide new tools for theoretical and computational chemistry.

Chemical systems are ultimately governed by the laws of quantum mechanics, which can be exceedingly difficult to simulate using computers that are ultimately only classical systems. It is estimated that quantum systems consisting of as few as even 50 quantum bits will not be able to be simulated by the world's most powerful computers. However, the creation of precise, controllable quantum systems which can mimic the dynamics of chemical systems is already rapidly approaching scalability. Harnessing the power of these quantum computers will likely be the key to unlocking new approaches to computational chemistry that were impossible to achieve in the past.

Despite the potential power of these systems, there is still much to explore in quantum information, a field which has only risen to prominence in the last two decades. The logic of quantum theory operates so differently from the classical intuition of the past that not many examples exist to exploit this power. Most quantum algorithms rely on only a handful of theoretical ideas to achieve computational speedups. Armed with only these, however, the intuition gained from quantum algorithms have at times given rise to new approaches for their classical counterparts as well.

Quantum algorithms that do have dramatic runtime improvements over classical algorithms so far require more quantum resources, in the form of qubits or number of qubit operations, with more precision and control than is currently available. While progress in existing platforms may realize quantum architectures outperforming classical ones over a time scale of some years, there remains potential for feedback from better experimental methods in chemistry to improve these techniques and to create novel platforms which shorten this time scale significantly.

Experimental progress in quantum information has been moving rapidly on many platforms, each with their own strengths. Quantum annealing machines now have thousands of interacting manufactured qubits. Impressive experimental control has now been demonstrated for small arrays of superconducting and ion trap qubits. Quantum optical devices can now be fabricated on-chip with on-demand single photon sources coming online in the next few years. Bose-Einstein condensates are routinely prepared in a wide variety of systems and long-range coherent ordered phases have been observed at room temperature. Our improving understanding and control of excitations in materials is opening new possibilities to engineer more efficient methods to transport energy and information both across nanodevices and between cities. Quantum topological effects have been observed in materials and are expected to yield superior devices for the storage of quantum information and the transport of energy.

> ...there is still much to explore in quantum information, a field which has only risen to prominence in the last two decades.

All of these platforms have potential for addressing different problems in chemistry, just as each can take advantage of new developments in chemical systems to improve their scalability and control.

Thus, a unique interface exists between chemistry and quantum information which enables progress in either of these fields to exponentially compound, making this a ripe opportunity to exploit. It was the goal of this workshop to identify the areas where cross fertilization between these two fields would result in the largest payoff for future developments in theory, algorithms, and experimental techniques.

## 1.3 WORKSHOP OVERVIEW

With support from the National Science Foundation, the organizers held a day and a half workshop on November 13 and 14th, 2016. The workshop had 43 participants that included a diverse group of faculty members, national lab employees and industry representatives, as well as graduate student and postdoctoral researchers as facilitators. Attendees included observers from several funding agencies.

The goal of the workshop was to summarize recent progress in research at the interface of quantum information science and chemistry as well as to discuss the promising research challenges and opportunities in the field.

The workshop was run as a collective brainstorming exercise featuring no talks other than short summaries of subgroup discussions. The organizers employed participation techniques similar to those employed by the Launch.org conferences. The collective brainstorming resulted in a set of raw Google documents comprising ~48 pages.

The ideas can be broadly categorized in two distinct areas of research that obviously have interactions and are not separated cleanly. The first area is **quantum information for chemistry**, or how quantum information tools, both experimental and theoretical can aid in our understanding of a wide range of problems pertaining to chemistry. The second area is **chemistry for quantum information**, which aims to discuss the several aspects where research in the chemical sciences can aid progress in quantum information science and technology.

A unique interface exists between chemistry and quantum information... making this a ripe opportunity to exploit.

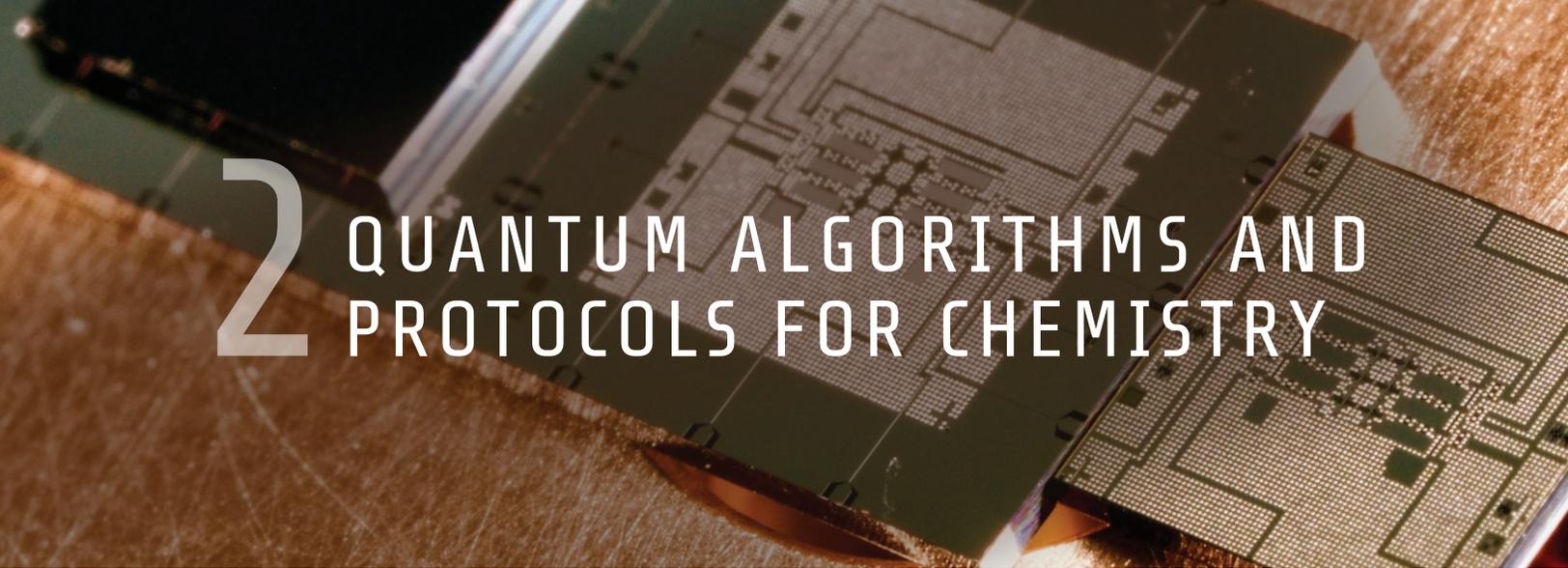

# 2 QUANTUM ALGORITHMS AND PROTOCOLS FOR CHEMISTRY

Chemistry is one area in which quantum computers could deliver a practical advantage over their classical counterparts

Among the prominent applications of quantum information processing, chemistry is one area in which quantum computers could deliver a practical advantage over their classical counterparts in the near term. In Section 2.1, we describe the development of quantum algorithms for simulating chemical systems and discuss prospects for advancing this field through further algorithmic development. Section 2.2 discusses using the current quantum information processors in conjunction with classical computers to improve the speed or accuracy of chemical calculations. Section 2.3 expands on how chemistry can provide the ideal arena for yielding insights on the deep scientific question "What makes quantum computers more powerful than their classical counterparts?". In Section 2.4, we explore the current lines of inquiry that can combine the powerful tools developed in both quantum computing and machine learning to advance chemistry even further than each one area has on its own.

## 2.1 QUANTUM ALGORITHMS FOR THE SIMULATION OF CHEMICAL SYSTEMS

The field of quantum information gained traction in the 1990s through the development of several landmark quantum algorithms, including Shor's algorithm for quantum factoring (Shor, 1999), and Grover's algorithm for searching in an unstructured database (Grover, 1996). In parallel, Feynman's proposal for simulating quantum systems (Feynman, 1986) was given a solid foundation through Lloyd's quantum simulation algorithms, which employ the techniques of Trotter-based quantum evolution (Lloyd, 1996) and quantum phase estimation (Abrams and Lloyd, 1999)**.** Building from these tools, Aspuru-Guzik et al. developed a quantum algorithm for computing molecular electronic ground state energies in polynomial time, paving the way for a variety of subsequent developments in quantum computing for chemistry (Aspuru-Guzik et al., 2005). In particular, this work sparked the development of quantum algorithms for the required task of quantum state preparation (Kassal et al., 2008), and similar approaches which handle multi-reference states (Wang et al., 2008). Researchers have also considered extending the use of quantum algorithms to various regimes in quantum chemistry such as relativistic dynamics (Veis et al., 2012) and quantum dynamics beyond the Born-Oppenheimer approximation (Kassal et al., 2008; Welch et al., 2014; Kivlichan et al., 2016). Error correction for quantum algorithms for quantum chemistry has also been investigated (Jones et al., 2012).

A crucial aspect of developing quantum algorithms for quantum chemistry is to estimate the resources required for performing relevant simulation tasks. Initial efforts in this regard have yielded resource estimates in the context of molecular simulation (Whitfield et al., 2011). The following years saw a sequence of developments in analyzing the asymptotic scaling of resources needed for performing quantum chemical simulations on a quantum computer, in terms of the numbers of electrons and spin-orbitals. In spite of seemingly pessimistic initial estimates (Wecker et al., 2014a), developments in recent years have led to significant improvements in resource estimates for simulating quantum chemistry (Hastings et al., 2015; Poulin et al., 2015; Babbush et al., 2015a). Recently, Reiher and co-workers carried out a detailed study of the computational cost, including quantum error correction, for FeMoCo, a model of the nitrogenase enzyme (Reiher et al., 2016) that suggests that it is indeed feasible to employ future error-corrected architectures for the simulation of realistic chemical systems of scientific and industrial interest.

The introduction of algorithms for the simulation of sparse Hamiltonians (Berry et al., 2015a; Berry et al., 2015b) made for a breakthrough in the field of quantum simulation. In contrast to Lloyd's original work, which used the Trotter-Suzuki approximation to simulate evolution under a Hamiltonian, Berry et al. instead considered the possibility of decomposing the Hamiltonian as a linear combination of unitary operators, and employing a truncated Taylor series to simulate Hamiltonian dynamics. These algorithms have been recently extended to several different contexts, including quantum chemistry simulation in both first (Babbush et al., 2015b) and second quantization (Babbush et al., 2016). These sparsity-based algorithms scale better asymptotically as a function of molecule size than all prior algorithms for quantum simulation.

It is indeed feasible to employ future error-corrected architectures for the simulation of realistic chemical systems of scientific and industrial interest.

Consideration for simulating quantum systems in the first-quantized form began in the 1990s with proposals for discretizing the wave function in real space, and constructing a similarly discretized Hamiltonian to simulate evolution under (Wiesner, 1996; Zalka, 1998). Subsequent works have opened up an entire line of inquiry concerning real-space simulation of quantum chemistry (Kassal et al., 2008; Welch et al., 2014; Kivlichan et al., 2016). In particular, it has been found that for systems with more than roughly 4 atoms, real space simulation is both more efficient and more accurate than second-quantized methods (Kassal et al., 2008; Kivlichan et al., 2016). These ideas have been combined with the recently developed sparse Hamiltonian simulation techniques (Berry et al., 2015a; Berry, et al., 2015b) to develop a real space simulation algorithm (Kivlichan et al., 2016) which outperforms its first- and second-quantized counterparts (Babbush, et al., 2015b; Babbush, et al., 2016) in certain regimes.

The results discussed so far are based on gate model quantum computers, which can be broadly categorized as the "digital" approach to quantum simulation, since the overall unitary quantum evolution being simulated is eventually decomposed into a sequence of elementary operations called quantum gates in a way similar to how logic gates underlie classical computers. However, there is an alternative "analog" approach to quantum computing based on adiabatic quantum evolution (Farhi et al., 2000) that has also gained much attention in the quantum information community. Rapid experimental progress has been underway to scale up devices for realizing adiabatic quantum computation (D-Wave Systems Inc., 2017). In connection with quantum chemistry, researchers have considered protein folding proposals for adiabatic quantum computers (Perdomo-Ortiz et al., 2012; Babbush et al., 2014b). Molecular electronic structure problems have also been mapped to an adiabatic model of quantum computation that involves additional coupling terms in the Hamiltonian (Babbush et al., 2014a; Cao et al., 2015).

As early quantum computers have become available, researchers around the world have employed them to carry out quantum chemistry calculations. In particular, few-qubit simulations, often without error correction, have been carried out in most major architectures used for quantum information. In 2010, Lanyon et al. demonstrated the use

**Few-qubit simulations, often without error correction, have been carried out in most major architectures used for quantum information.**

of the iterative phase estimation algorithm (IPEA) to measure the energy of molecular wavefunctions. In this case, the wavefunction of molecular hydrogen ($H_2$) in a minimal basis set was encoded in a 1 qubit state and the IPEA was realized using a two-qubit photonic chip (Lanyon et al., 2010). A similar procedure was applied to $H_2$ using nuclear magnetic resonance (Du et al., 2010) and to helium hydride ($HeH^+$) using nitrogen vacancies in diamond (Wang et al., 2015). More recently, experimental efforts have been focused on the quantum variational eigensolver (VQE) algorithm (see next section), the first demonstration of which was performed in 2014 (Peruzzo et al., 2014). This experiment employed a two-qubit photonic chip to variationally minimize the energy of $HeH^+$. A VQE experiment of the same molecule was achieved using a system comprising a single trapped ion (Shen et al., 2017). These experiments served as proofs of principle for the IPEA and VQE approaches applied to chemistry, but involved non-scalable simplifications of the Hamiltonians or non-scalable experimental procedures such as full tomography. The first scalable demonstration of both the IPEA and VQE algorithms employed three superconducting qubits for simulating $H_2$ in a minimal basis and was carried out by a Google Research and Harvard collaboration (O'Malley et al., 2016). These demonstrations were followed by the demonstration of VQE with a hardware-based ansatz for to $H_2$, lithium hydride (LiH) and beryllium hydride ($BeH_2$) in a system with six qubits (Kandala et al., 2017) by the IBM quantum computing team.

The field of quantum algorithms for electronic structure is still ripe for further development. Concepts such as wavefunction locality (McClean et al., 2014) can be introduced to further reduce the cost in terms of quantum gates for molecular simulation. The combination of these ideas with sparse algorithms and a smart choice of basis functions may reduce the cost of quantum simulation of chemical systems further. There are several ideas from the domain of classical molecular electronic structure than can be applied in the field of quantum simulation for further reduction of quantum computational cost. As reliable quantum computers begin to come online, there will be a continued demand for improvement of these quantum algorithms.

As reliable quantum computers begin to come online, there will be a continued demand for improvement of these quantum algorithms.

## 2.2 HYBRID QUANTUM-CLASSICAL ALGORITHMS

As quantum devices continue to develop, advantageous coupling between these devices and existing classical resources can help boost the potential of these new technologies. Hybrid quantum-classical (HQC) computation combines the strengths of quantum and classical computation, utilizing each where appropriate, as illustrated in Figure 1.

**FIGURE 1. HYBRID QUANTUM-CLASSICAL SCHEME FOR COMPUTATION.**

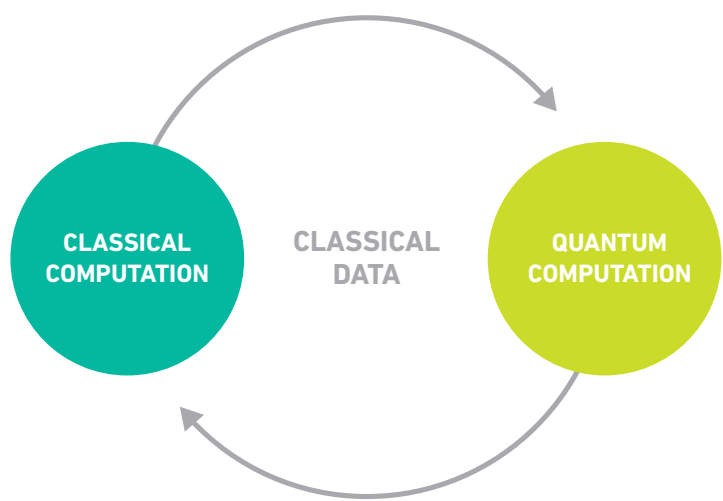

*Hybrid quantum-classical computing utilizes both classical and quantum hardware, sharing classical data between them, to try to exploit "the best of both worlds" for each computing model.*

Hybrid strategies utilize classical central processing units (CPUs) which select classically hard parts of an algorithm, and feed these instances into quantum processing units (QPUs) to solve. Conversely, more involved classical computation can reduce the overhead for an otherwise fully quantum computation. These algorithms are particularly relevant to the current era of quantum computing, known as the pre-threshold era. We have devices that are coherent enough to carry out non-trivial calculations but that are yet to achieve the threshold precision and number of qubits to realize full error correction. In the pre-threshold era, algorithms of low circuit depth, i.e. involving few quantum gates, are what is practical. Often, these algorithms are asymptotically less efficient than algorithms involving more complex procedures, and hence longer circuit depths, but in turn, are realizable nowadays. After error correction is achieved, HQC algorithms are still expected to be important for solving some classes of problems as they will achieve larger circuit depths.

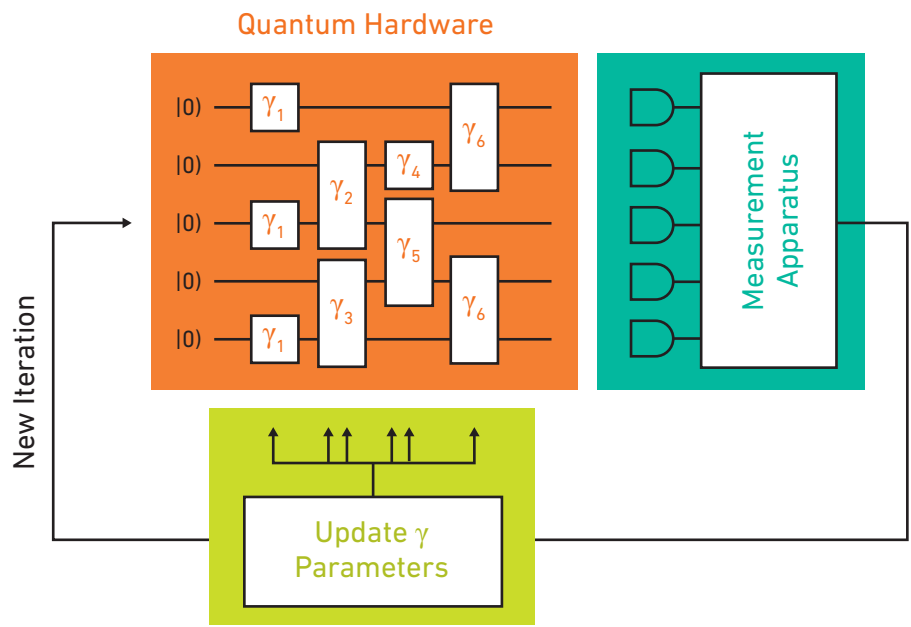

*Preparation and propagation of the quantum state occurs within the quantum hardware framework (or QPU), while the optimization of the parameters of the quantum state is achieved on the CPU by minimizing an objective function (Guerreschi and Smelyanskiy, 2017). The same scheme applies to other algorithms such as quantum autoencoders.*

Two representative HQC algorithms are the variational quantum eigensolver (VQE) (Peruzzo et al., 2013; Wecker et al., 2015; McClean et al., 2016; O'Malley et al., 2016; Santagati et al., 2016; Guerreschi and Smelyanskiy, 2017; McClean et al., 2017; Romero et al., 2017; Shen et al., 2017) and the quantum approximate optimization algorithm (QAOA), (Farhi et al., 2014a; Farhi et al., 2014b; Lin et al., 2014; Farhi et al., 2016) the workflows of which are both illustrated in Figure 2. The VQE allows for the simulation of the ground and excited states of otherwise-intractable quantum systems. It parametrizes a quantum state through some polynomial number of classical parameters, uses a quantum device to evaluate the expectation value of an objective function which depends on them in some way, and updates the parameterization through a classical non-linear minimization routine. VQE has been demonstrated experimentally and applied to the simulation of molecules using a variety of experimental platforms including quantum optics and superconductors. The QAOA has shown promise as an algorithm for approximately solving combinatorial optimization problems, a tool that is very useful for (but not limited to) chemistry.

The field of HQC algorithms has advanced in a variety of theoretical directions, including work on state parametrization, cost reductions and optimization strategies (Yung et al., 2014; McClean et al., 2016; Wecker et al., 2015; Yang et al., 2016; Romero et al., 2017; Guerreschi and Smelyanskiy, 2017). In addition, the VQE algorithm has been demonstrated for small molecules (O'Malley et al., 2016; Shen et al., 2017; Kandala et al., 2017), indicating that the method offers robustness to systematic experimental errors compared to phase estimation (O'Malley et al., 2016). Most recently, this method has been extended to include excited states using a quantum linear-response subspace method (McClean et al., 2017) and a combination of optimization of a purity metric together with quantum phase estimation (Santagati et al., 2016). The linear-response subspace method has the additional benefit of correcting some errors under decoherence quantum channels. Recent efforts have also enabled the use of these methods in thermodynamic systems by leveraging embedding techniques such as dynamical mean-field theory or density matrix embedding theory (Bauer et al., 2016, Dallaire-Demers and Wilhelm, 2016a; Dallaire-Demers and Wilhelm, 2016b; Rubin, 2016). The coupling of quantum and classical resources has also been studied with tools of classical resource analysis (Britt et al., 2015; Humble et al., 2016). These efforts help to quantify requirements for both HQC algorithms and the classical resources that will be required for future, fault-tolerant implementations of quantum devices.

HQC algorithms inspired by machine learning techniques have been also proposed. Specifically, a modified version of QAOA with a different objective function was employed to train quantum circuits that predict instances of the MAX-2-SAT problem (Wecker et al., 2016). More recently, a general purpose algorithm inspired by classical autoencoders (Bengio, 2009) was proposed for the compression of quantum information. This quantum autoencoder model (Romero et al., 2016) trains a quantum circuit to compress a set of quantum states, reducing the number of qubits required for storage. This method has potential applications in quantum simulation and quantum communication and could be employed to discover new state preparation protocols.

In the last several years, industrial and public efforts to engineer a medium-scale quantum computer based on superconducting qubits have made rapid progress (Barends et al., 2014; Kelly et al., 2015; Barends et al., 2015; Córcoles et al., 2015). It is expected that this effort will deliver pre-threshold quantum computers in the next couple of years. These machines, possessing approximately 50-100 qubits, will be able to execute 1000 to 10000 gates and will be likely accessed via a cloud system (Mohseni et al., 2017; Castelvecch, 2017). Although pre-threshold quantum computers will lack error cor-

> It is expected that [the effort to engineer superconducting qubits] will deliver pre-threshold quantum computers in the next couple of years.

rection, they will be able to perform tasks beyond the reach of the current classical supercomputers, achieving so-called quantum supremacy (Boixo et al., 2016). Workshop participants identified HQC algorithms as the best suited for pre-threshold quantum computers due their more efficient use of coherence time and their robustness against systematic experimental errors. However, there are several challenges that need to be overcome before HQC algorithms can be implemented in near-term hardware.

To experimentally demonstrate these HQC algorithms, we will need to consider the constraints posed by the hardware which may include restrictions in connectivity and implementable gate sets. These aspects will impact the final circuit depth and the number of measurements required for execution. One strategy that could help the incorporation of these variables in quantum algorithm design is the development of general purpose software that compiles high level quantum circuits all the way to hardware. Software packages such as ProjectQ (Steiger et al., 2016) have made this one aspect of their main priorities. In the context of quantum simulation of fermions, the FermiLib library for general purpose manipulation of fermionic and qubit operators has been recently launched and developed as a library used in conjunction with ProjectQ. An alternative strategy, particularly suited for VQE and QAOA algorithms, is to develop new circuit ansatzes tailored to specific hardware specifications (Peruzzo et al., 2013; Farhi et al., 2017; Kandala et al., 2017).

Apart from the ability to compile HQC algorithms to quantum hardware, software packages for quantum applications should be complemented with simulation tools. These simulators will accelerate the evaluation and testing of new HQC algorithms, allowing the validation of results obtained with quantum hardware for small to medium systems (<40 qubits). Several advances have been made towards the design of such tools (Wecker and Svore, 2014b; Smelyanskiy et al., 2016; Steiger et al., 2016) but there is still room for improvement. Challenges in this direction are the consolidation of current software packages still under development, the design of special purpose parallel computing techniques for quantum circuit simulation, the development of new simulation algorithms based on tensor network approaches and the incorporation of realistic noise models into simulations (Sawaya et al., 2016).

Although pre-threshold quantum computers will lack error correction, they will be able to perform tasks beyond the reach of the current classical supercomputers, achieving so-called quantum supremacy.

## 2.3 IDENTIFYING THE BOUNDARIES BETWEEN CLASSICAL AND QUANTUM COMPUTATION

As realizations of quantum computers grow in scale, an important research direction is to identify problems for which quantum algorithms have a practical advantage over the best known classical algorithms. The workshop participants have broadly identified efforts in the quantum information community so far as empirical and formal approaches. The empirical line of research focuses on estimating the physical resources needed for solving representative instances of a certain problem and compare the resource estimates for both quantum and classical algorithms. The formal approach, however, seeks to rigorously characterize whether a problem is efficiently solvable by a quantum or classical computer or is likely beyond efficient computation in the worst case, be it either quantum or classical.

Both empirical and formal approaches have delivered important insights on the boundaries between classical and quantum computation, and chemistry is the ideal arena in which these insights can be of both theoretical and practical value. This is because chemistry has a truly non-trivial stake in algorithmic performance - each time additional computational resources are made available, more chemical calculations that were previously out of reach become accessible (Barney, 2016). In this context, the answer to the question whether quantum mechanical resources of a quantum computer are required for accurate computation of molecular electronic properties is then also highly relevant. If the answer is affirmative, then this makes a perfect practical case for quantum computing. Otherwise, if we can show that quantum chemistry may be described classically in spite of its quantum nature, this can open the door to efficient exact solutions of these problems on a classical computer. This would motivate the devotion of significant national resources to achieve this transformative advance. Regardless of which way the question is resolved, the chemistry community stands to benefit, gaining a tool for simulating the electronic structure of molecules. The difference lies only in whether the simulations are performed on classical or quantum processors.

In the context of quantum chemistry (Aspuru-Guzik et al., 2005; Kassal et al., 2008; Whitfield et al., 2011; Kais, Dinner and Rice, 2014), identifying the boundaries between quantum and classical algorithms may also be considered along empirical and formal lines. On the empirical side, concrete resource estimates in terms of *e.g.* running time and qubit cost have been carried out for typical quantum chemistry simulation problems

> Chemistry is the ideal arena in which [insights from quantum information] can be of both theoretical and practical value.

on a quantum computer (Whitfield et al., 2011; Jones, et al., 2012; Wecker et al., 2014a; Poulin et al., 2015; Hastings et al., 2015). The connection between molecular features and the cost of quantum simulation has also been observed (Babbush et al., 2015a). These resource estimates are crucial first steps for a fair comparison with classical algorithms solving the same sets of quantum chemistry problems. From the formal perspective, tools from theoretical computer science have deepened our understanding of what makes a quantum system hard to simulate on a classical computer (Gharibian et al., 2015). These tools have also been used on quantum chemistry problems such as electronic structure calculation (Whitfield, Love, and Aspuru-Guzik, 2013) and density functional theory (Schuch and Verstraete, 2009) to provide a new perspective on the extent to which quantum computers may help current quantum chemical calculations. However, an outstanding line of research is to understand the implications of formal worst-case complexity analysis in chemistry in contrast to average case complexity of instances of chemical interest.

Although the present discussion follows the dichotomy of empirical and formal approaches, one might also consider a mixture of both methodologies. Empirical analyses are valuable on a case-by-case basis but lacks the generality that formal results deliver. Formal methods, on the other hand, are typically restricted to worst-case scenarios while the actual probability that such worst-case scenarios occur in practice remains to be assessed. One way to strike a balance between the two approaches can be seen in the recent study of "empirical hardness" of canonical problems that are widely believed to be hard to solve on classical computers (Leyton-Brown et al., 2014). It would be interesting (and perhaps of greater impact) to carry out similar methodologies to study the "empirical hardness" of quantum chemical problems. But whichever method is chosen, empirical, formal or in between, identifying the boundary between quantum and classical computers in solving quantum chemical problems will remain a subject of tremendous theoretical and practical relevance.

Tools from theoretical computer science have deepened our understanding of what makes a quantum system hard to simulate on a classical computer.

## 2.4 QUANTUM MACHINE LEARNING

The last decade and a half has witnessed an explosion of machine learning techniques – deep learning in particular – to analyze big data sets. The combination of large training sets and fast processors has given rise to a `virtuous cycle' of the development of ever more powerful and effective machine learning techniques that have had a significant recent impact on various areas of science such as particle physics (Racah et al., 2016; Aurisano et al., 2016; Renner et al., 2016; Acciari et al., 2017). The intuition that quantum mechanics could help with pattern recognition is straightforward: quantum systems are well-known to generate patterns that are hard to generate classically. Perhaps, then, they can also recognize patterns that are hard to recognize classically.

Classical machine learning has already been applied successfully to large chemical data sets. An example of this is the design of functional molecules stemming from screening databases (Gómez-Bombarelli et al., 2016). How can quantum machine learning allow us to learn patterns in chemical data that we cannot learn classically? The first point to note is that chemical wave function data, by its very nature, comes out of particular quantum mechanical systems whose statistical properties are difficult to generate classically. Quantum simulators can generate these patterns efficiently, and quantum machine learning techniques may find patterns in such data that classical machine learning techniques simply cannot.

In many cases, we can use quantum random access memory to store classical data more efficiently by taking advantage of quantum superposition (Giovannetti, Lloyd, and Maccone, 2008). Such compressed storage, combined with known quantum algorithms for basic linear algebra operations (Harrow et al., 2009), allowed for an entire generation of quantum machine learning algorithms to be developed (Wiebe et al., 2012; Rebentrost et al., 2014; Zhao et al., 2015; Schuld et al., 2016; Lloyd et al., 2016). Most of these new quantum machine learning techniques are exponentially faster than their classical counterparts, under some assumptions.

The continued development of quantum machine learning algorithms as well as their implementation in near-term hardware can be expected to increase our capabilities in classifying and predicting chemical data, leading to accelerated molecular design timescales. Some of the challenges identified in the workshop include the possibility of developing novel quantum machine learning algorithms specifically tailored to chemistry – to learn electronic structure, probe the interactions of molecules with light, and find reaction pathways, amongst other chemical applications.

The continued development of quantum machine learning algorithms... can be expected to increase our capabilities in classifying and predicting chemical data.

Another important direction for chemistry in the context of quantum machine learning is considering the quantum realization of neural networks. Most quantum neural network construction in the literature so far has been able to take advantage of the unique features of quantum mechanics, but their inner workings remain rather distinct from classical neural networks and none of the proposals so far have captured all the essential features of classical neural networks (Schuld et al., 2014). It is then crucial to bridge the gap between the burgeoning communities of both classical and quantum neural networks, so that one may be able to leverage the insights yielded in the vast literature of classical deep learning community for improving quantum neural networks. Particularly for chemistry, recent years have seen rapid progress in using classical neural networks for solving problems such as molecular design and screening (Duvenaud et al., 2015; Gómez-Bombarelli et al., 2016). Being able to use quantum computers to realize classical neural networks with improved performance could then in turn augment their applications in chemistry.

Compared with simulating neural networks for classical data on quantum computers, more developed is the line of inquiry on quantum neural networks for *quantum* data, namely quantum states. Progress has been made in using quantum generalizations of classical neural networks for learning essential features of a quantum state in a way that classical computers cannot (Kieferova and Wiebe, 2016). Moreover, ideas inspired by constructions of classical neural networks, such as autoencoders (Hinton and Salatudnikov, 2006), have already reaped some initial success in compressing quantum wavefunctions (Romero et al., 2016). The coming years will see a flurry of progress being made in developing quantum machine learning algorithms for extracting information from quantum states. Among other applications, chemistry will be one of the first areas to benefit from the progress in this direction.

Being able to use quantum computers to realize classical neural networks with improved performance could then in turn augment their applications in chemistry.

# 3 QUANTUM PLATFORMS FOR CHEMISTRY

Quantum algorithms with features that are ideal for optical networks have been devised for the calculation of important quantities in Chemistry.

The large variety of problems in chemistry requires a diverse set of theoretical and computational methods to solve. In line with these problems, many quantum platforms were identified by the workshop participants, which leverage their system to tackle one or more of these challenges specifically. In Section 3.1, we address quantum optical systems that can take advantage of optical networks to perform phase estimation and directly sample from distributions corresponding to molecular spectra. In Section 3.2 we discuss how strong coupling can control and modify molecular energy pathways, merging ideas from quantum optics with single-molecule quantum chemistry. We present recent work in the fast-moving field of quantum electrodynamical (QED) chemistry and the challenges yet to be overcome. Building on this, in Section 3.3, we explore the possibility of using optical cavities to monitor and control the kinetic and thermodynamic properties of chemical reactions. In Section 3.4, we highlight the tremendous potential of single molecule polaritons with strong nonlinearities with wide ranging applications from quantum information and logic to modifying the thermodynamic and kinetic behavior of chemical reactions for new synthetic and spectroscopic methods. In Section 3.5, we explore the possibility of using polariton condensates to improve the coherent control of reaction pathways and phase transitions. Beyond condensates, Section 3.6 presents concepts for fermionic simulators based on optical lattices of ultracold atoms to solve quantum many-body problems. Finally, in Section 3.7, we discuss the use of multidimensional optical spectroscopy ('quantum spectroscopy') that uses the quantum behavior of photons, unlike classical spectroscopy, to characterize molecules in a fundamentally different way.

## 3.1 QUANTUM OPTICAL TOOLS FOR CHEMISTRY

The fields of atomic and molecular optics (AMO) and quantum information have worked together to establish systems with very large system-cavity couplings to carry out quantum information tasks. The workshop participants identified several directions where cross-fertilization amongst fields could result in breakthroughs.

Currently, quantum algorithms with features that are ideal for quantum optical networks have been devised to carry out computation of important quantities in chemistry. In particular, an optical implementation of the protocol to compute Franck-Condon profiles (Huh et al., 2017) is quickly nearing reality, reinforced by its similarity to Boson Sampling (Wang et al., 2016). These profiles are able to produce the spectrum of vibronic frequencies of molecules undergoing electronic transitions. Classical computation of this quantity would require solving a computational problem which is notoriously hard, the matrix permanent. However, quantum optical networks like the one in Figure 3 have the capability to sample from a distribution with probabilities proportional to the square of the permanent without needing to compute this quantity directly, producing the profile far more efficiently than by direct computation.

**FIGURE 3. PICTORIAL DESCRIPTION OF BOSON SAMPLING AND MOLECULAR VIBRONIC SPECTROSCOPY.**

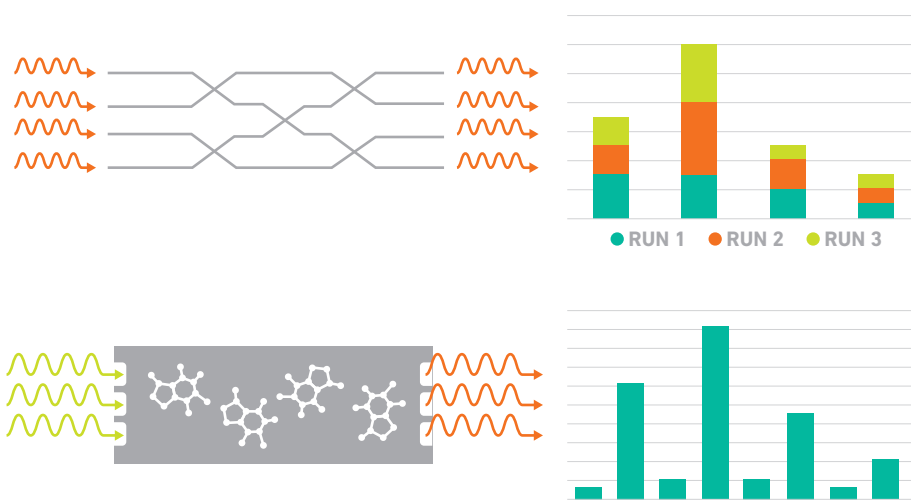

**a.** Boson sampling consists of sampling the output distribution of photons obtained from quantum interference inside a linear quantum optical network. **b.** Vibronic spectroscopy uses coherent light to electronically excite an ensemble of identical molecules and measures the re-emitted (or scattered) radiation to infer the vibrational spectrum of the molecule. The fundamental physical process underlying **b** is formally equivalent to situation **a** together with a nonlinear state preparation step.

Quantum optical sensors which exploit entanglement on a few-photon level are capable of performing super-sensitive and super-resolving measurements (Hosten et al., 2016; Motes et al., 2015; Olson et al., 2017). Although scaling up these sensors to outperform classical devices on macroscopic and stable objects would be very challenging, for biological and chemical systems that are sensitive to light, there is a natural application. As these systems can be perturbed or even destroyed by repeated probing, experiments are limited in the number of measurements that can be safely applied. Maximizing the precision and resolution for these measurements is crucial to learning more about the processes and properties of these materials.

Previously, the quantum information community has focused its efforts on generating particular quantum states of light that maximize their sensitivity, but have not typically addressed the topic of signal-to-damage ratio (Juffmann et al., 2016). Efforts that drive research in this direction may also lead to understanding more of the properties of the quantum states of light (such as sensitivity to various sources of noise, such as loss or dephasing) that lend themselves to better light-matter coupling for qubits or other chemical systems.

## 3.2 CONTROLLING AND PROBING MOLECULAR ENERGY TRANSFER PATHWAYS VIA STRONG-COUPLING

In the strong-coupling regime hybrid quasi-particles of novel character, with mixed light-matter character, can dramatically change the chemical landscape. In the new field of quantum electrodynamical (QED) chemistry, for example, chemical reactions are performed in optical high-Q cavities without the need to explicitly drive the system externally. Hybridization of electronic states with the optical mode in strongly coupled molecule-cavity systems has revealed unique properties: lasing (Kéna-Cohen and Forrest, 2010), room temperature polariton condensation and the modification of excited electronic landscapes involved in molecular isomerization.

A key idea in advancing our knowledge of the behavior of complex molecular systems in the strong coupling regime has recently been explored in the study of strong exciton-photon coupling between molecular complexes and a confined optical mode within a metallic optical nanocavity. Picocavities have also been demonstrated in recent work (Benz et al., 2016). Of special relevance to our discussion are recent experimental studies that have explored the strong coupling regime using organic dye molecules in cavities of

*Hybrid quasi-particles of novel character, with mixed light-matter character, can dramatically change the chemical landscape.*

relatively small Q. These low Q cavities, compared to the ones employed for atoms, are substantially easier to fabricate. This fundamental fabrication step forward is possible due to the fact that dyes have much larger transition moments and could be exploited in spectroscopic studies of molecules (Coles et al., 2014a; Coles et al., 2014b; Chikkaraddy et al., 2016; Zhang et al., 2012).

The possibility of modifying chemical reaction rates (i.e. kinetics) has been suggested based on several earlier experiments where bulk properties were modified by strong coupling, such as the work-function and the ground-state energy (Vasa et al., 2013). However, changing the energy landscape and energy transfer pathways via strong coupling remains a challenge. Theoretical and computational descriptions of such systems go beyond the well-established concepts in quantum chemistry and quantum optics and necessitate new methods developed for the field of QED chemistry (Galego et al., 2015; Herrera and Spano, 2016; Flick et al., 2017; Martínez-Martínez et. al., 2017). Progress in quantum electrodynamical density-functional theory and Cavity Born-Oppenheimer approximation methods could play a key role in the prediction (Frediani and Sundholm, 2015; Herrera and Spano, 2016) and design of novel experiments, as identified in the workshop.

## 3.3 CONTROLLING AND MONITORING CHEMICAL REACTIONS USING OPTICAL CAVITIES

A sizable number of experiments and theoretical studies have shown that the chemical properties and reactivity of molecules can be significantly modified when molecules interact strongly with cavities (Galego et al., 2015; Kowalewski et al., 2016a; Kowalewski et al., 2016b; Csehi et al., 2016; Casey et al., 2016; Muallem et al., 2016; Martínez-Martínez et. al., 2017). This interaction has been also exploited to design new spectroscopic techniques (Shabani et. al, 2014) to study reaction dynamics (Matsugi et al., 2016). The participants of the workshop identified the use of optical cavities to control chemical reactions as an exciting field of research that is rather unexplored. Future work in this direction should focus on determining the limits of control of chemical thermodynamics and kinetics that can be achieved for molecule-cavity systems, both experimentally and theoretically. Some of the challenges in the theory side are the development of tech-

Changing the energy landscape and energy transfer pathways via strong coupling remains a challenge.

niques for describing situations where the Born-Oppenheimer approximation fails as well as the development of electronic structure methods for studying molecules in the strong-coupling regime (Bennett et al., 2016; Flick et al., 2017). While traditional methods from quantum chemistry, such as the Born-Oppenheimer (BO) approximation, Hartree-Fock theory, coupled-cluster theory or density-functional theory (DFT) are able to describe the quantum mechanical nature of the states in matter, they cannot account for the quantized nature of the electromagnetic field. On the other hand, conventional methods from quantum optics typically explore the quantum electromagnetic field in great detail (Forn-Diaz et al., 2016), but ignore the 'matter' component (it is mostly described by simplified models restricted to only a few-energy levels). Both over-simplifications, in quantum chemistry, and quantum optics, are far from a realistic description of QED chemistry, where many atoms and molecules each with electronic and nuclear degrees of freedom are coupled strongly to the vacuum electromagnetic field. In this few-photon limit, the quantized nature of the electromagnetic field is a key aspect that has to be considered to correctly describe vacuum fluctuations, spontaneous emission, polariton-bound states or thermalization. New theoretical studies would aim at explaining some of these new intriguing features observed in strong-coupling or even ultrastrong-coupling (Martínez-Martínez et. al., 2017) experiments.

**FIGURE 4.**

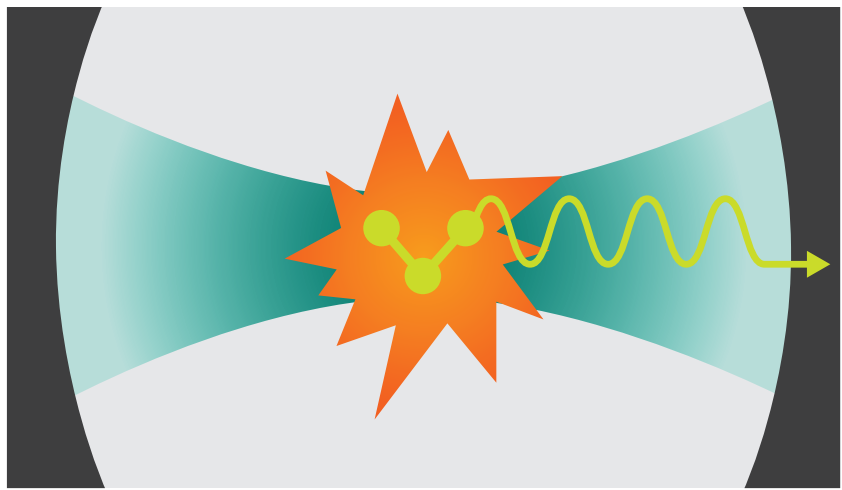

*Optical cavities could modify the kinetic and thermodynamic properties of chemical reactions, opening the path to new synthetic methods allowed by these 'nanofurnaces'. In addition, optical cavities could be used to develop new spectroscopic methods for monitoring reaction dynamics.*

## 3.4 SINGLE-MOLECULE POLARITONS IN 'PICOCAVITIES'

Recent experimental advances have shown that individual atomic features inside the gap of a plasmonic nanoassembly (cavity) can localize light to volumes well below 1 cubic nanometer, forming 'picocavities' that enable optical experiments and single-molecule probes on the atomic scale (Chikkaraddy et al., 2016; Benz et al., 2016). These single molecule polaritons (SMPs) are achieved by strong coupling between a molecular exciton and a tightly confined electromagnetic field in a plasmonic nanostructure (Narang et al., 2016; Mertens et al., 2017). So far, it has been shown that these strong optical field gradients can switch the Raman selection rules and the ultrasmall localization of light in these cavities alters the number and variety of vibrational modes of trapped molecules observed. This experimental breakthrough offers tremendous possibilities for the control of potential energy surfaces of molecules, and therefore, the adiabatic and non-adiabatic dynamics in photochemical processes. From a different point of view, SMPs offer strong nonlinearities that can be exploited for nanophotonic and quantum information logic applications.

## 3.5 POLARITON CONDENSATES

Bose-Einstein condensation (BEC) of molecular polaritons (MP) can occur at room temperature owing to the strong binding energies of organic excitons as well as the very strong light-matter coupling afforded in microcavities. Can exotic collective quantum phenomena, previously reserved for ultracold matter, be harnessed for chemical applications? It turns out that this is indeed a possibility, where the many-body effects of condensates guide chemical reactions with almost perfect selectivity towards a target reaction (Moore and Vardi, 2002).

BEC is the process whereby the ground state of a bosonic system becomes macro-scopically occupied, giving rise to nontrivial many-body effects such as superfluidity (Landau, 1941). Heuristically the onset of BEC occurs when the de Broglie $\lambda_D = \sqrt{\frac{h}{(2\pi m k_b T)}}$ wavelength of the quasiparticle in question becomes larger than the characteristic interparticle separation, which depends on the quasiparticle density. In three dimension, the critical temperature for non-interacting particles is given by $n\lambda_D^3 = 2.62$. There are two possibilities to observe a crossover to BEC from a normal gas: to lower the critical temperature to cryogenic conditions or to work with low mass quasiparticles. Tradition-ally, ultracold temperatures are needed to reach the onset of atomic gas BECs. On the other hand, owing to their partial photonic character, the effective mass of polaritons can be a very small fraction of the mass of the electron, thus allowing for the possibility that is at room-temperature (Byrnes et al., 2014). Even though this effective mass argu-

This...
breakthrough
offers tremendous
possibilities
for the control of
potential energy
surfaces of
molecules.

ment is independent of whether we are discussing inorganic polaritons or MPs, it is only correct for the latter given that the small binding energies of inorganic excitons precludes polariton formation at room temperature, thus still requiring cryogenic temperatures for the onset of BEC (Kasprzak et al., 2006). On the other hand, MP BEC has been recently predicted (Bittner and Silva, 2012), experimentally demonstrated (Galbiati et al., 2012; Plumhof et al., 2014) and shown to exhibit superfluidity.

The possibility of room temperature BEC for molecular systems opens doors to novel considerations for the chemical sciences. A particularly intriguing opportunity is to harness BEC for coherent control of reactive processes, such as originally proposed in (Moore and Vardi, 2002). The scheme relies on amplification mechanisms enabled by boson scattering. Suppose that a reactant can branch into two products, which can be labeled as $A$ and $B$. Let the coupling amplitude between the reactant and each of the products be $\Omega_A$ and $\Omega_B$. Moore and Vardi have predicted that if such a reactant starts in BEC form with an average of particles, the time evolution of the branching ratio will be $N_B(t)/N_A(t) \approx Exp[\sqrt{N(\Omega_B - \Omega_A)t}]$, where $N_i(t)$ is the population of each product. Thus, nearly uniform branching ratios can be deformed into near to perfectly selective transformations upon use of a BEC.

It remains to be explored whether this mechanism still holds for MP BECs in the presence of delocalized polariton states, nonequilibrium conditions, and specific photochemical conditions. This enhancement mechanism is strongly dependent on the macroscopic (average) population of the "reservoir/source" state. This implies that population fluctuations are small enough relative to $N$, such that they can be completely ignored. Even though it is tempting to mimic this result with a multimode room-temperature lasing system, the latter does not exhibit this collective quantum mechanical behavior, since it does not feature a single macroscopically occupied quantum state in the gain medium. In this objective, we aim to exploit the collective dynamics of BECs in order to drive kinetic processes with high selectivity. These include selective chemical reactions and phase transitions.

Just as BECs can selectively drive a chemical reaction, we suspect that they can do the same with phase transitions. Generically, at a given set of macroscopic conditions (e.g., temperature, pressure, chemical potential, etc.) a material behaves in equilibrium according to the phase which minimizes its free energy. However, metastable/transient phases may be achieved under non-equilibrium or quasi-equilibrium conditions. These may also be studied experimentally if sufficiently long-lived. It has been found in the past twenty years that metastable phases of a variety of solids can be obtained under electronic excitation by irradiation with light. Examples include transformations from ionic to covalent, or from high to low spin phases in organic charge-transfer complexes

**Just as Bose–Einstein condensates can selectively drive a chemical reaction, we suspect that they can do the same with phase transitions.**

and conjugated polymers. These are generally explained to arise due to the presence of strong electron-phonon coupling in one or more electronic excited states, that is, in cases where the lattice relaxation is strong enough to induce the formation of precursor domains corresponding to phases that are inaccessible under thermodynamic equilibrium conditions. The rate of formation of the new phase is proportional to the density of nucleation domains. Given that the metastable phase is thermodynamically favored in a given excited state, the density of precursors is expected to be strongly enhanced if the excitation of the solid induces formation of a Bose–Einstein condensate-like state. Just as with their reactive counterparts, it is not yet clear whether the materials from which non-equilibrium polariton BECs can be formed are the same as those for which photoinduced phase transitions exist. However, it is tenable, e.g., there exist photoinduced phase transitions which have been observed to happen on a few femtoseconds, i.e., quick enough for a hypothetical polariton BECs not to decay before. Furthermore, if the rate is accelerated by the rate of BEC, the transition will happen even faster. In summary, polariton BECs could facilitate the formation of metastable solid-state phases by enhancing the rate of formation of nucleation domains. The employment of BECs in this field would open the door to the discovery of novel phases suppressed in thermodynamic equilibrium. Interest in the discovery of such exotic phases lies in their potentially useful new properties. While they have an obvious drawback due to their relaxation to the equilibrium phase under sufficiently long time, chemical doping as well as other strategies in materials design can be envisioned as ways to overcome this issue. We propose to theoretically study the amplification of nucleation domains for metastable photoinduced phases driven by polariton BECs.

**FIGURE 5. CONTROLLING REACTION RATES WITH POLARITON CONDENSATES**

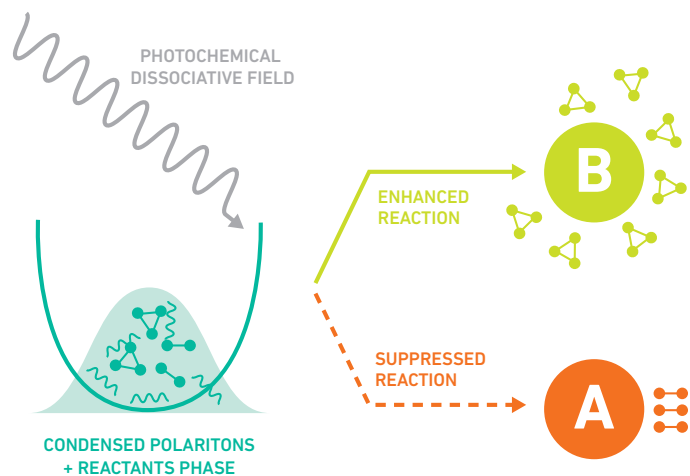

*Starting from a condensate phase, a dissociation path is favored by a photochemical dissociative field.*

## 3.6 FERMIONIC SIMULATORS FOR ELECTRONIC STRUCTURE

Optical lattices are periodic standing waves of light, created by interfering optical laser beams. These lattices have been widely used as potential energy landscapes to cool, trap and control neutral atoms as shown in Figure 6. Recent experimental advances have allowed the placement of single atoms on each lattice site, the imaging and manipulation at the single-site level, and an exquisite level of control of their interactions, even at the single-site or tweezer level. All these features have turned optical lattices of ultracold atoms into one of the most promising systems for quantum simulation of many-body systems.

Simulations using optical lattices are now performed for strongly correlated systems in condensed-matter physics such as Hubbard models (Bloch et al., 2014), topological matter (Goldman et al., 2016) and orbital physics (Li and Liu, 2016). However, current approaches fail to capture the physics of more complex systems such as molecules. As mentioned in the introduction of this report, the exact simulation of the electronic structure of molecules on a classical computer is thought to require an exponential number of resources relative to the system size and therefore is limited to a few atoms in classical computers. An alternative that exhibits only polynomial scaling are digital quantum computers, composed of qubits. Because qubits do not follow fermionic statistics, simulating fermions with qubits requires mapping procedures that introduce an extra level of complexity in the simulation. The use of optical lattices for atoms or molecules opens the possibility of creating quantum simulators that employ *indistinguishable* spin ½ particles or fermions to carry out the simulation.

The major challenge for the simulation of arbitrary molecules is engineering arbitrary 3D potentials.

During the workshop, we identified two possible approaches for building a *fermionic simulator* for molecules. In the first approach, atoms in an optical lattice are employed to directly emulate the behavior of molecules, which we call the "potential-based approach." In this case, ultracold atoms mimic electrons while the optical trapping potential is used to mimic the attractive potential produced by nuclei (Lühmann et al., 2015). So far, this approach has been applied to planar artificial molecules, that emulate the behavior of real systems such as benzene. The major challenge for the simulation of arbitrary molecules is engineering arbitrary 3D potentials. This could be achieved by using three-dimensional nonseparable lattice potentials formed by several interfering laser beams (Nogrette et al., 2014) or by holographic projection techniques, as suggested by (Lühmann et al., 2015).

It may eventually be possible to introduce a degree of freedom for simulating time-dependent chemical processes as well. Because the lattice potential can be dynamically changed (Tarruell et al., 2012), i.e. because it can be made time-dependent, important chemical processes such as chemical reactions, isomerization, and photodissociation may be simulatable. Yet another degree of freedom can be introduced to the optical lattice by using a mixture of ion types (Kohstall et al., 2012). There may be scenarios in which a Hamiltonian can be mapped more efficiently to such a mixture rather than to a homogenous set of fermionic ions.

The second proposal for achieving fermionic simulation can be identified as an "operation sequence approach," where dynamical control sequences (i.e. a set of operations performed over time) executed on ultracold atoms would allow for the simulation of molecular fermionic interaction terms. This proposal has never been realized for molecules and could open the door to a novel method for studying electronic structure using systems that have been traditionally employed to simulate only homogeneous Hamiltonians in the field of condensed-matter physics.

The Hamiltonian of molecular systems can be written in second quantization as a sum of hopping terms, $a_p^\dagger a_q$ that capture the nuclear-electron interaction as well as the kinetic energy of electrons, and terms acting on four orbitals, $a_p^\dagger a_q^\dagger a_r a_s$, that account for the electron-electron repulsion. In the operation sequence simulation approach, atoms mimic the electrons in a molecule, and each spin-orbital corresponds to a location in the *optical lattice*, as illustrated in Figure 6. The electrons in a molecule can either hop between spin-orbitals or interact with each other. In this fermionic simulator, hopping corresponds to moving atoms from one lattice site to *any* other, and electronic interaction corresponds to an interaction between atoms in up to four *different* lattice sites. The implementation of these interaction terms could be achieved using a stroboscopic technique that properly alternates the atoms' hopping and interactions, effectively implementing the electronic structure Hamiltonian. The major challenge for the realization of this approach is the implementation of long-range hopping and repulsion interactions between arbitrary sites of the optical lattice.

The implementation of *long-range hopping* between arbitrary lattice sites could be realized by adding a wide trapping potential overlapping with all lattice sites, and making both the optical lattice and the wide trap *state-dependent* trapping potentials. This means that atoms in an internal ground state are only trapped by the lattice, and atoms in an internal excited state are only trapped by the wide trap. Two internal ground and excited states of an atom such as strontium can be selected such that they are connected by an optical transition. This would allow for the transfer of an atom from one site to another via the wide trap using a laser tuned to the appropriate frequency. Another possibility is to use a mobile optical tweezer instead of the wide trap, as recently demonstrated with rubidium (Kaufman et al., 2015). After setting up a deformable micromirror device to address single lattice sites, long-range hopping can be implemented. *Long-range interactions* could be realized by implementing an additional wide trapping potential connecting another pair of lattice sites. The interactions on the wide traps can be realized either via optical or orbital Feschbach resonances (Saha et al., 2014; Höfer et al., 2015).

**FIGURE 6. ILLUSTRATION OF THE A) "POTENTIAL BASED" AND B) "OPERATION SEQUENCE" APPROACHES FOR THE SIMULATION OF A MOLECULAR HAMILTONIAN WITH A FERMIONIC SIMULATOR.**

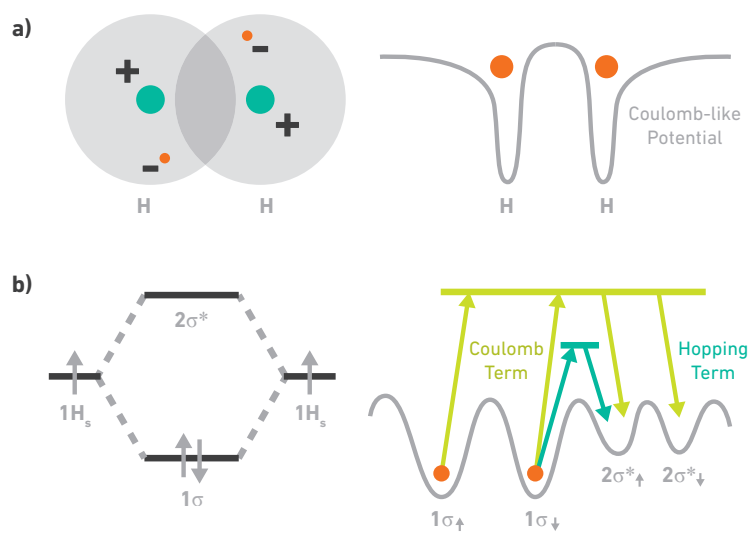

*The example shows the mapping of a hydrogen molecule in a minimal basis set to elements in an optical lattice. In the potential-based simulation, the potential landscape of the optical lattice mimics the Coulomb potential generated by nuclei and ultracold atoms play the role of electrons. In the operation sequence approach, the sites of the optical lattice represent orbitals of the molecule and the hopping and coulomb (electron–electron repulsion) interactions present in the molecular Hamiltonian are effected on the ultracold atoms.*

The simulation of a molecule could be achieved by adiabatically evolving the Hamiltonian starting on a suitable initial state such as the one obtained from a Hartree-Fock (HF) ansatz. The Hamiltonian terms would be implemented at each stroboscopic cycle. The final state produced by the simulation will be inferred using a quantum gas microscope, which will image the quantum state and number of atoms on each lattice site at the end of the simulation. Finally, the development of fermionic simulators will promote advances in the theory of fermionic quantum computers. These machines offer other advantages with respect to qubit based computers apart from the natural treatment of fermionic systems, such as reduced sequences for gate implementation, as shown by (Bravyi and Kitaev, 2002).

## 3.7 MULTIDIMENSIONAL OPTICAL SPECTROSCOPY FOR QUANTUM INFORMATION PROCESSING

The response of complex molecular systems to sequences of ultrafast optical laser pulses can provide a multidimensional view of the quantized electronic and vibrational degrees of freedom. Amenable systems include the unpaired electron spin in diamond nitrogen vacancy (NV) centers, the delocalized electronic states of coupled molecular networks, and the electronic spins of electronically excited charge-transfer states in molecules. The development of new spectroscopic techniques may increase the viability of using such systems for quantum information processing or quantum sensing. Conversely, use of theoretical concepts from quantum information, such as quantum process tomography, continue to give new life to techniques that enable the unravelling of the dynamics of quantum systems (Yuen-Zhou et al., 2011; Yuen-Zhou et al., 2014a; Yuen-Zhou et al., 2014b).

Conventional nonlinear spectroscopy uses classical light to probe properties of matter through the variation of its response with pulse frequencies or time delays. New "quantum spectroscopy" techniques have paved the way for major advances in characterization and control of molecular quantum systems. Quantum spectroscopy offers new control parameters and enables increased signal strength at low photon fluxes by exploiting quantum properties of light such as coherence and photon entanglement. Entanglement may be employed to improve the precision of measurement beyond the

Quantum process tomography, continue[s] to give new life to techniques that enable the unravelling of the dynamics of quantum systems.

Heisenberg limit and enhance spatial resolution in quantum imaging and quantum lithographic applications, as well as in quantum-optical coherence tomography. Recent experiments demonstrated the control of exciton distributions, and the suppression of exciton transport in photosynthetic complexes thanks to the combination of high temporal and spectral resolution offered by entangled light.

Towards quantum information applications, advances in quantum spectroscopy must be geared towards addressing single molecular quantum systems having few degrees of freedom. Meeting this objective requires the design and development of new experimental techniques involving novel laser pulse sequences, high sensitivity, and the associated theoretical and computational tools for their analysis.

As for progress in this direction, a stimulated Raman technique has been proposed (Dorfman, 2014), which combines entangled photons with interference detection to select matter pathways and enhance the resolution. Following photoexcitation by an actinic pump, the measurement uses a pair of broad-band entangled photons; one (signal) interacts with the molecule and together with a third narrow-band pulse induces the Raman process. The other (idler) photon provides a reference for the coincidence measurement. This interferometric photon coincidence counting detection allows to separately measure the Raman gain and loss signals, which is not possible with conventional probe transmission detection. This setup can better resolve fast excited-state dynamics than possible by classical and correlated disentangled states of light.

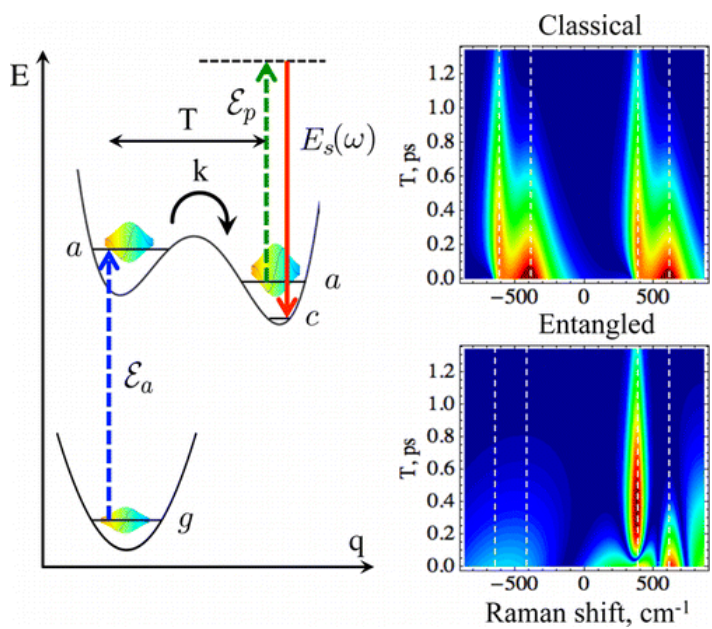

Ultrafast quantum process tomography protocols have been implemented for chemistry for the first time experimentally by using a transient grating technique on a double-walled J-aggregate sample (Yuen-Zhou et al., 2014). The prospects of carrying out more sophisticated spectroscopies that employ non-classical light and continue using concepts such as quantum process tomography is an opportunity to advance our understanding of the dynamics of chemical systems at the ultimate detail possible.

Entanglement may be employed to improve the precision of measurement ... and enhance spatial resolution in quantum imaging and quantum lithographic applications, as well as in quantum–optical coherence tomography.

# 4 CHEMISTRY FOR QUANTUM INFORMATION

Chemistry is fundamental to the development of complex matter platforms that are relevant for many applications ranging from energy production and storage, healthcare, and now, quantum information processing. The potential uses of chemistry for building novel quantum information processing elements from the bottom-up was heavily discussed in the workshop. Chemistry is also crucial for optimizing existing materials for the current quantum information processing platforms.

## 4.1 BOTTOM-UP CHEMICAL QUBITS

A common discussion theme was the use of the arsenal of synthetic chemistry tools for developing new methods to develop atomic- and molecular-scale qubits. Molecular chemistry may lead to the development of designer quantum molecules with the favorable quantum properties comparable to those of diamond color centers. While spin and optical properties of these molecular quantum systems may be slightly diminished because of stronger interactions with the environment, achieving structural control at the atomic level would provide a major upside, leading to arrays of quantum memories or ultra-small quantum sensors for chemistry, biology, and material science. 2D architectures in particular offer a stable platform for designing such materials, and recent experiments have revealed some of the brightest room-temperature quantum emitters reported in such materials. We therefore identify a major opportunity for chemists to develop, through theory and experiment, mid-scale chemical systems with well-controlled spin, optical, and mechanical properties.

> Achieving structural control at the atomic level would provide a major upside, leading to arrays of quantum memories or ultra-small quantum sensors for chemistry, biology, and material science.

Existing photonic networks are "passive", meaning that they use only linear optical elements such as beam-splitters, mirrors and so on. Combined with single photon measurements and feed-forward of these measurement outcomes these linear optical networks are capable of universal quantum computation. Even without feed-forward such networks can instantiate hard sampling problems, Boson Sampling being the pre-eminent example. However, "active" networks with nonlinearities have the ability to carry out universal quantum computation with much lower overhead. While passive, linear optical, components can now be carried out on an all-optical chip, nonlinearities at the single photon level are currently hard to implement. Atoms and Molecules are natural nonlinear elements. Designer molecules capable of realizing improved optical nonlinearities on-chip would enable the possibility of implementing sophisticated quantum algorithms on a large scale quantum photonic platform.

Below, we review the promising chemical classes discussed in the workshop. In all of these classes, computational materials discovery is poised to play an effective role. In the past decade, algorithmic searches for new types of materials have led to significant discoveries of materials that are governed by quantum physics, including photovoltaics, batteries, and photocatalysts (Jain et al., 2017). As materials informatics matures as a field, and as machine learning methods emerge for predicting quantum behavior, these tools will be directly applicable to the discovery and design of materials perfectly suitable for quantum information processing.

**Multi-spin Organic Molecules and Assemblies:** Organic molecules provide an inexpensive and highly tailorable platform on which to develop systems for quantum information processing and quantum computing (Boehme and Lupton, 2013; Sun et al., 2014). One key advantage that organic molecules have over most of their inorganic counterparts is that the electrons in organic molecules generally have weaker spin-orbit couplings, which results in longer electron spin coherence lifetimes. This means that spin information can be more easily maintained and manipulated (Brown et al., 2011; Wesenberg et al., 2009; Sato et al., 2009) or transported (Kandrashkin and Salikhov, 2010; Salikhov et al., 2007; Kobr et al., 2012; Krzyaniak et al., 2015) through complex molecular assemblies. The scalability of ensemble and single molecule electron spin-based systems for quantum computing and information has been widely recognized.

It is important to note that chemical synthesis makes it possible to design molecules in which the spin-spin interactions between multiple electrons and nuclei can be controlled by the chemical bonds that link them. Moreover, the future holds the possibility that

These tools will be directly applicable to the discovery and design of materials perfectly suitable for quantum information processing.

molecular recognition and self-assembly strategies can be used to tailor these interactions with the ultimate goal of self-assembling quantum computing and information systems from simple molecules.

Covalently-linked electron donor-acceptor (D-A) molecules are one category of systems having intriguing possibilities for use in quantum computing and information systems. Fast photo-initiated electron transfer within a D-A molecule can result in the formation of an entangled pair of electron spins with a well-defined initial spin configuration (Hashamoni et al., 1995; Miura and Wasielewski, 2011). Time-resolved electron paramagnetic resonance (TREPR) and pulsed EPR spectroscopies provide important means of manipulating and controlling these coherent spin states (Schweiger and Jeschke, 2001). These radical pair states in D-A systems have displayed coherent spin motion for ~100 ns, (Hoff et al, 1998; Kothe et al., 1994; Laukenmann et al., 1995) which could provide a basis for the development of new quantum computing and information processing strategies based on electron spin coherence. Before these strategies can be exploited, a greater understanding of the factors controlling spin coherence dephasing is necessary. For example, it has been demonstrated that the spin coherence generated within a photo-generated radical pair can be transferred to a second radical pair using laser-driven electron transfer, while preserving spin coherence for tens of nanoseconds at room temperature (Kobr et al., 2012; Krzyaniak et al., 2015).

Photoexcitation of organic chromophores often generates highly electron spin-polarized triplet states, which can transfer this polarization to nuclear spins (Kothe et al., 2010; Lin et al., 2005). Application of pulsed electron-nuclear double resonance (ENDOR) spectroscopy has been shown to provide a method of transferring spin coherence and polarization from the electrons to the nuclei, and has been used to implement a rudimentary quantum logic gate. More extensive work has been done on organic molecules containing multiple radicals in which the spin-spin interactions allow the implementation of quantum logic gates, (Sato et al. 2009; Yamamoto et al., 2015) which are manipulated using pulsed ENDOR.

Computation most often requires information transfer between processors and memory, so that in most models of quantum computation this transfer must remain coherent at all times (Mehring et al., 2003; Simmons et al., 2011). A major requirement for any physical qubit is the preparation of a pure initial state. One way to approach this in an electron spin system is by generating high initial spin polarization on a single electron spin. Until now, most quantum information processing experiments using electron spins have been performed in the weak spin polarization regime. Achieving high thermal electron spin

> Computation most often requires information transfer between processors and memory, so that ... this transfer must remain coherent at all times.

polarization requires applying high magnetic fields and very low temperatures to an isolated spin system. Creating spin memories requires manipulating qubits on a variety of timescales and moving information to and from more environmentally isolated states. Nuclear spins benefit from longer relaxation times than electron spins, but suffer from intrinsically weak thermal polarization, resulting in impure initial nuclear spin states with the attendant difficulties in manipulating these spin states. Once again, using pulsed ENDOR techniques, it has been demonstrated that spin coherence transfer from an electron spin to a nuclear spin and back again can extend the coherence time of the qubit to more than one second (Morton et al., 2008). Another class of qubits that could potentially be designed consists of emergent topological defects in vibronic systems such as conical intersections in Jahn-Teller molecules (Ribeiro and Yuen-Zhou, 2017).

**Metal-organic frameworks (MOFs) and Covalent-organic frameworks (COFs):** One key advantage conferred by coordination chemistry is the well established ability to create arrays of individual units. Supramolecular chemistry is nearing the point where it will enable the design of complex architectures where each precise atomic position can be configured. More recently, the expansion of metal-organic frameworks and covalent organic frameworks brought in a solid-state-like structural rigidity to the creation of coordination materials. Here, the modularity of these systems generates precisely spaced qubit units. Further structural control is imparted by the potential for postsynthetic modification, where individual units can be exchanged. This capability may enable rational design of long-lived qubits within these solid-state architectures. For example, if a specific phonon process is implicated in decoherence, a postsynthetic modification could enable swapping out the ligand of concern.

Once synthesized, depending on the dimensionality of the MOF, it is possible to either exfoliate the MOF to create a 2D surface, in the case of a 2D MOF. Otherwise, electrode-position, both cathodic and anodic, demonstrated significant strides towards creating surfaces. Both of these approaches could be refined as a mechanism of creating reproducible arrays of qubits.

MOFs and COFs can form an astoundingly large variety of structures, belonging to a range of space groups. This may allow for the design of qubits arrays with novel underlying lattice structures, beyond standard two-dimensional arrangements like square and kagome lattices, to include nearly any 3D regular lattice type. Because certain quantum information processing paradigms, including topological quantum computation, require a particular spatial arrangement of qubit units, MOFs and COFs may become the materials of choice for certain applications.

Supramolecular chemistry is nearing the point where it will enable the design of complex architectures where each precise atomic position can be configured.

MOFs also offer significant opportunities for sensing. Their permanent porosity enables their evacuation. Evacuated frameworks could be employed as biological sensors or sensors for local magnetic fields. There are many studies of the biocompatibility of metal-organic frameworks for their potential in drug delivery. Using the established biocompatible MOFs for biological sensing would enable a precise determination of the location of a qubit sensor relative to an analyte. For more fundamental scientific discovery, MOFs could be used to study the interaction between various analytes and qubits. For example, one could envision studying the changes in coherence time of a porous array of qubits as helium or hydrogen gas are diffused into the framework. Further, the qubits could probe the coherent processes within a MOF. There are numerous possibilities for discovery within these systems.

Avoiding interaction with the environment enables the realization of long coherence times. An alternate approach is to insulate qubits from the environment electronically by employing an analogous approach to that found in atomic clocks. In two systems, atomic clock-like transitions have been employed to enable long coherence times. Avoided level crossings create a small local region where the derivative of energy vs. magnetic field is zero. At those points the frequency of a given transition is stable to small changes in the local magnetic field, thereby stabilizing the transition to external interference.

**Conjugated oligomers and polymers with radical sites:** Photo-generated electron hole pairs and how they couple with covalently attached stable radicals may provide a new gating mechanism (Yu et al., 2016; Fataftah et al., 2016). Potential molecular architectures could include Donor-Acceptor dyads appended with stable organic radicals. (Chernick et al., 2006; Mi et al., 2006; Colvin et al., 2013) Photo-excitation of the donor results in electron transfer to the acceptor. The resultant electron hole pair may then be entangled with the stable radicals spins. Important fundamental studies would include the nature of the entanglement as a function of substitution pattern and number of attached radicals. Other aspects related to readout would be D-A-radical systems having large (measureable) magnetooptical activity in the excited state but not in the ground state (Kirk et al., to be submitted 2017).

New molecular architectures that combine electron delocalization and exchange coupling are needed. Such molecules, polymers or materials will be required to create important components of quantum information hardware. Such architectures will require conjugated pi-systems as well as stable organic radicals. The combination of organic mixed-valence species with organic radicals to make new "double-exchange" molecules (Kirk et al., 2009), and radical-substituted conjugated polymers and oligomers (Oyaizu and Nishide, 2009) are promising architectures.

Over the past 20 years there has been extensive interest in the preparation of multi-spin organic compounds such as molecules, dendrimers and polymers. Factors affecting the strength of exchange interactions have been elucidated, with stable high-spin species having been prepared (Gallagher et al., 2015). Many, if not most, organic radicals are redox-active, offering the possibility of switching or gating as well as qubit preparation. Very little has been done to explore the coherence times in radical based systems, including how coherence times are affected by exchange coupling with other spin centers and magnetic nuclei.

**Designer 2D materials:** Recently, bright and photostable quantum emitters were discovered based on atomic defects in 2D materials. These include the brightest room-temperature single photon emitters that are also spectrally tunable (Tran et al., 2016; Schröder et al., 2016; Grosso et al., 2016) and can be spectrally narrow. However, we are just scratching the surface: there is still uncertainty about the crystallographic origin of these quantum emitters, and the search for new emitters is ad-hoc. But while the crystallographic origin of quantum emitters is notoriously difficult to determine in bulk host materials, 2D materials actually present us with every atom on the surface, which in principle allows the precise structural determination using nanoscale imaging.

More broadly, 2D materials offer a completely different way of discovery of promising quantum defects. They allow an atoms-up design and fabrication, through theory (using for example density functional theory or other quantum chemistry simulations) of localized quantum systems with all of the properties that we desire. Compared to bulk material hosts, these 2D material hosts would allow for control of coupling to the phonon bath, spin bath, and could greatly facilitate electrical control (since we can make electrical contacts and apply top/bottom gates). The advantages are equally strong on the fabrication and materials side: unlike bulk hosts, 2D materials allow access to every atom (e.g., through scanning probe techniques). The ability to design electronic defects in 2D materials using quantum chemistry simulations, together with the capacity to experimentally produce these "trapped molecules", makes the 2D material platform extremely promising for atoms-up control, design, and scaling of quantum information processing and sensing systems. Getting there will require advanced theory and simulation tools, atom-scale fabrication techniques, and scaling to mid- and large-scale quantum systems.

Two dimensional materials offer a completely different way of discovery of promising quantum defects.

**Endohedral fullerenes:** One approach to mimic the solid-state environments found in silicon carbide and diamond is to employ nuclear spin free cage complexes such as fullerenes (Zhou et al. 2016). Since decoherence is inherently caused by interaction with the environment, by isolating qubits in within nuclear spin free spheres it is possible to prevent interaction with the environment. Until recently, the best molecular candidates for qubits were endohedral fullerenes, which isolate an electronic spin from the environment in a manner akin to the isolation provided by diamond for NV centers.

While these systems are highly promising, there are two significant challenges to developing them into viable candidates for quantum information processing . Each of these challenges stems from the complete environmental isolation of these spins. For sensing applications, fullerene's ability to isolate the electronic spin reduces its viability as a sensor. Further, fullerenes, are not maximally biocompatible. The second challenge towards implementation follows from the idea that scaling quantum computational systems necessitates contact between individual qubits. There, again the complete isolation of these systems poses a significant challenge. Finally, there remain synthetic hurdles to isolating these systems, but they would likely be surmountable if the other challenges could be overcome.

**Organometallic complexes (coordination compounds):** Until last year the only molecules that were competitive with electronic spin based solid-state materials were the nitrogen- and phosphorus-doped endohedral fullerenes. Considering the success of nitrogen-vacancy centers and fullerenes, these are both systems where an electronic spin is isolated within a nuclear spin free matrix. The matrix is the diamond in the case of the NV center, and the fullerene in the latter case. Applying the same design principles to coordination compounds enables a determination of the upper limit of coordination compounds, in a similar manner to the isolation of a single NV center and its measurement. Creating this comparable measurement necessitates isolating a coordination complex within a similar nuclear spin free matrix. Towards that end, Freedman and coworkers isolated a coordination complex in a nuclear spin free ligand field, and dissolved it in the nonpolar nuclear spin free solvent, carbon disulfide. (Zadrozny et al., 2015) They observed millisecond coherences times comparable to those observed in NV centers in diamond. This valuable proof of concept suggests the potential for employing arrays of

Creating this comparable measurement necessitates isolating a coordination complex within a similar nuclear spin free matrix.

coordination complex qubits, in the arrays specified in the MOFs discussed previously. Further, Freedman and Wasielewski recently demonstrated that this is a generalizable approach by measuring a series of vanadyl moieties in the exotic polar nuclear spin free solvent sulfur dioxide. Within this solvent, they observed three molecular species exhibiting coherence times of $T_2 > 100$ μs. (Yu et al., 2016)

**Molecular magnets:** Molecular magnets are crystalline materials that are composed of repeating molecular building blocks, with nonzero spin. They are a natural choice for quantum information processing, as quantum hardware can be thought of as a series of coupled molecules each corresponding to a different qubit (Troiani et al., 2005). Because they form regular repeating structures, molecular magnets offer the advantage of a large number of quantum states. Additionally, their non-degenerate transitions allow for operations with fewer errors (Leuenberger and Loss, 2001). As with the materials discussed above, there is a substantial chemical space to explore in improving this material class.

**Magnetic impurities in quantum dots:** There has been some research investigating the viability of magnetic impurities quantum dots, for use as qubits (Ochsenbein and Gamelin, 2011). There is a significant advantage of using quantum dots instead of bulk magnetic impurities: the qubits (magnetic impurities) can be more easily coupled to other subsystems, for example by placing an organic ligand with a particular nuclear spin close to the dot's surface. Additionally, coherence times of impurities in the quantum dots are significantly longer than those in bulk, for the ZnMnO material studied. Fortunately, there is an arsenal of synthetic methods for precisely tuning the microenvironment surrounding the quantum dots' magnetic impurities. These synthetic strategies include choosing organic surface ligands, doping or alloying the dot's primary material, and modifying the shape of the quantum dot's surface. Finally, quantum dots can be straightforwardly combined into colloidal crystals, which may be advantageous for creating a designed lattice of qubits.

...there is a substantial chemical space to explore in improving [molecular magnets].

**FIGURE 8. EXAMPLES OF CHEMISTRY PARADIGMS THAT HOLD POTENTIAL FOR IMPROVING MATERIALS USED IN QUANTUM INFORMATION SCIENCES.**

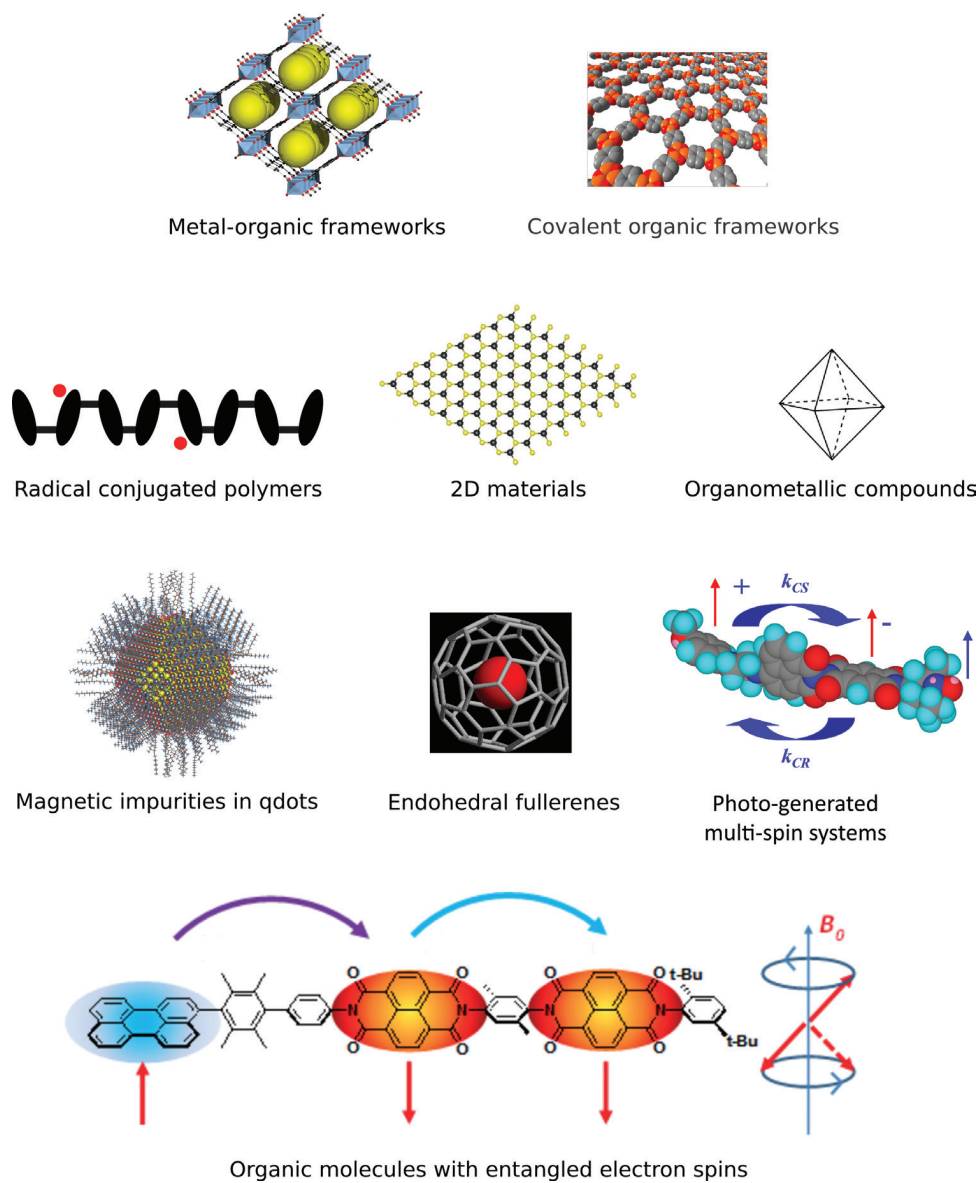

Metal-organic frameworks

Covalent organic frameworks

Radical conjugated polymers

2D materials

Organometallic compounds

Magnetic impurities in qdots

Endohedral fullerenes

Photo-generated multi-spin systems

Organic molecules with entangled electron spins

*Organic molecules with entangled electron spins.*

*Image attributions: MOF - Tony Boehle / Wikimedia Commons / Public Domain; COF - Boasap / Wikimedia Commons / CC-BY-SA-3.0 / GFDL; 2D material - 3113lan / Wikimedia Commons / CC-BY-SA-3.0 / GFDL; Qdot - Zherebetskyy / Wikimedia Commons / CC-BY-SA-3.0 / GFDL; Fullerene - Hajv01 / Wikimedia Commons / CC-BY-SA-3.0 / GFDL.*

## 4.2 ADVANCED QUANTUM READOUT TECHNIQUES

A crucial step in realizing molecular systems as quantum information processing units is the development of reliable readout techniques, necessary for quantum measurement and quantum error correction. Approaches to addressing this challenge must cater to the specific molecular quantum systems and various degrees of freedom (e.g. charge states, electronic excitations, electron- and nuclear-spin states) that are to be utilized.

Magnetic resonance-based techniques are best suited for measurement of electronic or nuclear spin-manifolds. Though long used, magnetic resonance approaches are inherently insensitive to measurements of single spin systems. Thus, efforts must be devoted to overcoming this insensitivity barrier, analogous to similar barriers surpassed in charge quantum dots, impurity spins of phosphorus in silicon and NV centers in diamond. A promising strategy to overcome this problem uses photo-driven formation of radical ion pairs to create highly spin polarized electrons; thus, greatly increasing the sensitivity of magnetic resonance readout techniques.

In comparison, readout of molecular quantum systems using optical detection (via ultraviolet through visible wavelengths) is rather reliable. Here, the remaining barrier is that, for many systems, this readout involves the destruction of the molecular system. The challenge, then, is to find molecular quantum systems which can withstand repeated encoding and readout in optically addressable degrees of freedom.

Finally, a very promising method for achieving readout in molecular systems is the use of scanning probe techniques. Over the past decade, a variety of scanning probe techniques for single-spin detection have been developed. These rely on either the direct detection of magnetic forces (Rugar et al., 2004), the utilization of NV centers for scanning probe controlled optical readout, or the utilization of spin-selection rules. In addition to possibly enabling quantum readout of molecular-based single quantum systems, scanning probe techniques may provide the only straightforward pathway for a random addressability of arrays of quantum systems.

## 4.3 CHEMICAL QUBITS FOR SENSING APPLICATIONS

Qubits sensors, such as NV centers, have been successfully employed to detect small variations in temperature, magnetic or electric field (Schirhagl et al., 2014). Qubit sensors measure static environmental properties such as magnetic fields and temperature by

The challenge, then, is to find molecular quantum systems which can withstand repeated encoding and readout in optically addressable degrees of freedom.

detecting changes in the qubit transition frequency. Oscillating environmental properties are easily detected if they are resonant with the qubit-frequency or if quantum control methods can be used to transform the oscillating signal into a static DC signal. However atomic and solid-state qubits have a limited set of transition frequencies and as a result can only detect a small fraction of the electromagnetic spectrum.

Another interesting application of qubit sensors is imaging. Luminescence from NV centers, for example, can be applied for imaging biological processes (Schirhagl et al., 2014). However, this application is limited by the compatibility of diamond nanoparticles with the biological environment as well as by the physical properties of the NV center. Specifically, NV sensors have low sensitivity to electric fields compared to magnetic fields, long measurement times and limitations in the tunability of relaxation times that prevent improvements in sensitivity of measurement.

To tackle these challenges, we propose the design of molecular qubits for sensors. The design space of molecular chemistry allows for complete control of the resonance frequency and the relaxation times. The polar nature of molecules makes them natural detectors of electric fields. In addition, the careful design of fluorophores would allow selective optomagnetic transitions tailored towards real time readout of the spin in complex environments. We also note that molecular sensors can be designed to enhance atomic and solid-state qubits. One example of this is the use of molecular scaffolding on the outside of a qubit containing nanocrystals to increase the versatility of NV centers. The same idea could be applied to quantum dots and other solid-state qubits.

One key advantage of nitrogen vacancy centers which is yet to be replicated in a molecular complex or material is their optical addressability. The ability to initialize and readout NV centers optically is primary to their success. This optical readout is enabled by the nature of the transitions within this system. NV centers are S = 1 compounds with a 3A ground state and a 3E excited state. The intersystem crossing and subsequent relaxation leads to the selective polarization of the Ms = 0 state. Transitions between the Ms = 0 and Ms = 1 states can be manipulated with microwaves, but the difference in the fluorescent energy between the two levels can be read out optically. If this seemingly simple system's success could be replicated within a molecular complex or material it would enable a revolution within sensing applications. The ability to attach a molecular qubit sensor to a biological sample and probe it through a chemical process would generate significant new fundamental knowledge about biochemistry.

Furthermore, one challenge with NV centers as probes is the lack of knowledge with regard to their precise location and orientation within a diamond. The synthetic biochemistry community possesses the well-honed capability to tether molecules to specific biological substrates and compounds. Achieving this goal would require designing a molecule with a comparable electronic character to an NV center. Coordination complexes offer synthetic flexibility towards that goal. Another key consideration is that while proof-of-concept species may be easier to achieve within a coordination complex, true realization of this concept would likely occur in a rigid structure such as a metal-organic framework (see Section 4.1). There the structural rigidity of the material would force the specific metal center to maintain a given geometry with distortion.

Another interesting research direction is the extension of the techniques employed in atomic clocks to molecular qubits. For example, the same quantum information techniques employed to make precise time measurement using the clock could be employed to enhance molecular ion qubit detectors (Schmidt et al., 2005). This would require the design of techniques to co-trap a molecular ion with an atomic ion qubit detector, something that could be achieved in the context of ultracold molecules (Carr et al., 2009).

Finally, molecular ion qubit sensors could be used to accurately measure temperatures, laser intensities (Wolf et al., 2016), and perhaps even changes in fundamental constants (Kajita et al., 2014). More directly, the laser-cooled atomic ions can be used to enhance molecular ion spectroscopy which has a wide application in astrochemistry (Khanyile et al., 2015), as shown in Figure 9. In addition there is a growing international community of chemists and physicists exploring the use of atomic ion qubits to study cold reactions (Chang et al., 2013).

**FIGURE 9. AS AN EXAMPLE OF HOW QUANTUM INFORMATION TECHNOLOGIES ENABLE CHEMISTRY, THE MOLECULAR SPECTRA OF *CaH*⁺ WAS OBTAINED USING LASER-COOLED ATOMIC IONS IN A CHAMBER BUILT FOR ION TRAP QUANTUM INFORMATION.**

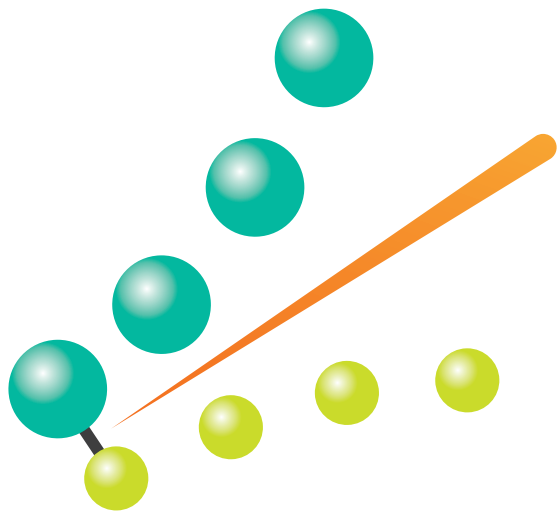

*The extremely sensitive nature of these quantum devices allowed the information to be obtain one molecular ion at a time. CaH⁺ is predicted to exist in the solar atmosphere but a lack of laboratory experiments has prevented*

> Molecular ion qubit sensors could be used to accurately measure temperatures, laser intensities and perhaps even changes in fundamental constants.

## 4.4 TOPOLOGICALLY PROTECTED MOLECULAR EXCITATIONS

A very promising approach to induce a fundamental protection against noise and dissipation in energy and information transportation and storage systems is to engineer topological effects in materials and nanodevices.

From the Aharonov-Bohm effect to the quantum Hall effect and exotic materials like topological insulators and superconductors, there is a deep connection between quantum mechanics and the abstract field of topology. A system with broken anti-unitary symmetries like time-reversal, charge conjugation or chirality may be endowed with non-trivial topological properties, where some of its states are protected against local perturbations and cannot be continuously deformed into one-another without non-local control parameters. A very promising approach to induce a fundamental protection against noise and dissipation in energy and information transportation and storage systems is to engineer topological effects in materials and nanodevices.

There are molecular electronic states that, owing to topological effects in their energetics, can move nanoscale energy unidirectionally and robustly even in the presence of disorder. These states have been proven to exist in simple excitonic molecular crystals as well as plasmonic metal-molecular aggregate molecular polaritons (Yuen-Zhou et al., 2014c; Yuen-Zhou et al., 2016), and could be exploited in light harvesting, molecular logic and nanophotonic applications.

Another approach to harness topological effects in chemistry for quantum applications would be to build up lattices from polaritonic molecules. Specifically, the Kitaev model (Kitaev, 2006) requires several functional components which could be incorporated through synthetic chemistry. Creating a honeycomb lattice of strongly coupled anisotropic metal centers could lead to a realization of these systems. Furthermore, chemically accessible oxidation states are also crucial to achieving a sufficiently stable compound for measurements. Therefore, the two most tractable metal centers are $Ru^{3+}$ and $Ir^{4+}$. $Ir^{4+}$ is a less desirable candidate due to its scarcity, cost, and tendency to reduce to form $IR^{3+}$. $Ru^{3+}$'s stability to reduction, large number of known precursors and extensively studied chemistry recommend it as a candidate for a Kitaev lattice. Recently the first experimental system posited to possess Kitaev physics was reported with a-$RuCl_3$ (Banerjee et al., 2016). The Neel ordering transition at 17°K, along with the complex stacked pattern of layers complicates the physics. Moving towards a molecular building block approach offers potential to build upon this exciting result.

## 4.5 THEORETICAL CHEMISTRY TOOLS FOR QUANTUM SIMULATION

The fields of computational chemistry and materials science have assembled a wealth of broadly-used theoretical techniques for nearly a century, for example the gold standard of coupled cluster theory or multireference methods for molecular electronic structure and quantum dynamics. Armed with these, researchers are able to convert physical questions about complex many-body quantum systems into tractable computational problems. Certain roadblocks faced by these classical approaches, such as the exponential growth of classical resources (Lloyd, 1996) and the dynamical sign problem (Gubernatis et al., 2001), are expected to be overcome by the use of quantum computers. Yet, just as with classical computation, quantum computation will face its own dichotomy between tractable and intractable physical problems. As small quantum information processors become available, there is growing demand to identify important problems which these devices are able solve and to develop theoretical approaches for bringing challenging problems into reach. While many of these future techniques will be uniquely suited to quantum computing, a great opportunity lies in building upon and repurposing theoretical tools from computational chemistry to convert complex quantum simulation problems into tractable ones. Difficult problems for quantum simulation include many-body spin systems with long-range correlations (Brandao and Kastoryano, 2016) and systems interacting with complex non-Markovian environments. Classical path integral techniques for mapping interacting many-body quantum systems into non-interacting quantum systems or for mapping non-commuting quantum problems into commuting ones may be useful for transforming these seemingly intractable quantum simulation problems into tractable ones.

A great opportunity lies in building upon and repurposing theoretical tools from computational chemistry to convert complex quantum simulation problems into tractable ones.

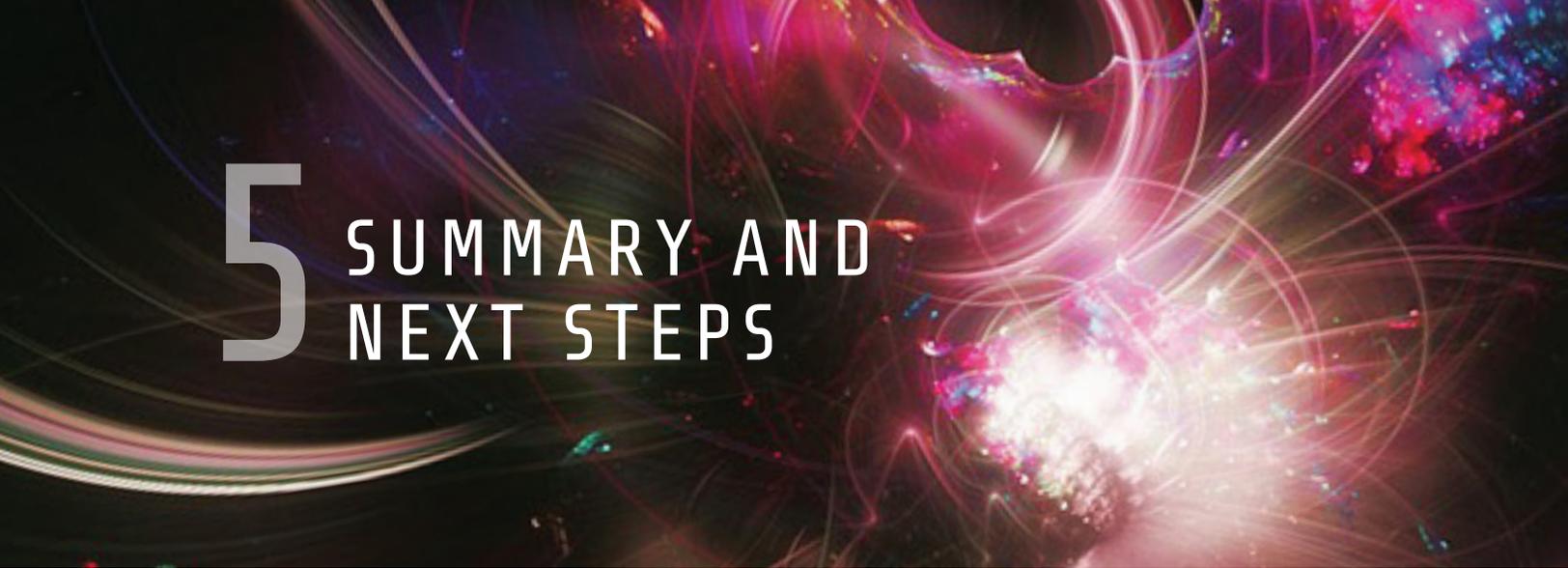

# 5 SUMMARY AND NEXT STEPS

Molecules and materials as inherently quantum mechanical entities can be studied using quantum information and also employed to carry out quantum information processing. The directions detailed above sample just a fraction of the scientific possibilities that a strong effort using chemistry-related ideas cross-fertilized with quantum information could open. Based on much of the work presented here, a review/perspective publication is in preparation to be submitted to a major Chemistry journal. The authors of this summary are open to inquiries while this process is underway.

# 6 GLOSSARY

| | |
|---|---|
| **AMO** | Atomic and molecular optics |
| **BEC** | Bose-Einstein condensation |
| **BO** | Born-Oppenheimer |
| **COF** | Covalent-organic framework |
| **CPU** | Central processing unit |
| **D-A** | Donor-acceptor |
| **DFT** | Density-functional theory |
| **ENDOR** | Electron-nuclear double resonance |
| **HF** | Hartree-Fock |
| **HQC** | Hybrid quantum-classical |
| **IPEA** | Iterative phase estimation algorithm |
| **MOF** | Metal-organic framework |
| **MP** | Molecular polaritons |
| **NV** | Nitrogen vacancy |
| **QAOA** | Quantum approximate optimization algorithm |
| **QED** | Quantum electrodynamical |
| **QPU** | Quantum processing unit |
| **TREPR** | Time-resolved electron paramagnetic resonance |
| **VQE** | Variational quantum eigensolver |

# 7 BIBLIOGRAPHY

Abrams, D.S. and S. Lloyd, "Quantum Algorithm Providing Exponential Speed Increase for Finding Eigenvalues and Eigenvectors," *Physical Review Letters* 83 (1999): 5162-165.

Acciarri, R., C. Adams, R. An, J. Asaadi, M. Auger, L. Bagby, B. Baller, et al., "Convolutional Neural Networks Applied to Neutrino Events in a Liquid Argon Time Projection Chamber," arXiv:1611.05531 (2017).

Aspuru-Guzik, A., Dutoi, A.D., Love, P.J. and M. Head-Gordon, "Simulated Quantum Computation of Molecular Energies," *Science* 309 (2005): 1704-1707.

Aurisano, A., A. Radovic, D. Rocco, A. Himmel, M. D. Messier, E. Niner, G. Pawloski, F. Psihas, A. Sousa, P. Vahle, "A Convolutional Neural Network Neutrino Event Classifier," *Journal of Instrumentation* 11 (2016): 09001.

Babbush, R., P.J. Love, and A. Aspuru-Guzik, "Adiabatic Quantum Simulation of Quantum Chemistry," *Scientific Reports* 4 (2014a).

Babbush, R., J. McClean, D. Wecker, A. Aspuru-Guzik, and N. Wiebe, "Chemical Basis of Trotter-Suzuki Errors in Quantum Chemistry Simulation," *Physical Review A* 91 (2015a): 022311.

Babbush, R., D.W. Berry, I.D. Kivlichan, A.Y. Wei, P.J. Love, and A. Aspuru-Guzik, "Exponentially more precise quantum simulation of fermions I: Quantum chemistry in second quantization," *New Journal of Physics* 18 (2016): 033032.

Babbush, R., D.W. Berry, I.D. Kivlichan, A.Y. Wei, P.J. Love, and A. Aspuru-Guzik, "Exponentially more precise quantum simulation of fermions II: Quantum chemistry in the CI matrix representation," arXiv:1506.01029 (2015b).

Babbush, R., A. Perdomo-Ortiz, B. O'Gorman, W. Macready, and A. Aspuru-Guzik, "Construction of Energy Functions for Lattice Heteropolymer Models: Efficient Encodings for Constraint Satisfaction Programming and Quantum Annealing," *Advances in Chemical Physics: Volume 155* (2014b): 201-44.

Banerjee, A., C.A. Bridges, J.-Q. Yan, A.A. Aczel, L. Li, M.B. Stone, G.E. Granroth, M.D. Lumsden, Y. Yiu, J. Knolle, S. Bhattacharjee, D.L. Kovrizhin, R. Moessner, D.A. Tennant, D.G. Mandrus, and S.E. Nagler, "Proximate Kitaev quantum spin liquid behaviour in a honeycomb magnet," *Nature Materials* 15 (2016): 733-740.

Barends, R., A. Shabani, L. Lamata, J. Kelly, A. Mezzacapo, U.L. Heras, R. Babbush et al., "Digitized adiabatic quantum computing with a superconducting circuit," *Nature* 534 (2016): 222-226.

Barends, R., J. Kelly, A. Megrant, A. Veitia, D. Sank, E. Jeffrey, T.C. White et al., "Superconducting quantum circuits at the surface code threshold for fault tolerance," *Nature* 508 (2014): 500-503.

Barney, L., "Modern Quantum Chemistry Research Possible Because of Supercomputers, HPC Software," *Scientific Computing*, (2016).

Bauer, B., D. Wecker, A.J. Millis, M.B. Hastings, and M. Troyer, "Hybrid quantum-classical approach to correlated materials," *Physical Review X* 6 (2016): 031045.

Bengio, Y., "Learning deep architectures for AI," *Foundations and Trends in Machine Learning* 2 (2009): 1-127.

Bennett, K., M. Kowalewski, and S. Mukamel, "Novel photochemistry of molecular polaritons in optical cavities," *Faraday Discussions* 194 (2016): 259-282.

Benz, F., M.K. Schmidt, A. Dreismann, R. Chikkaraddy, Y. Zhang, A. Demetriadou, C. Carnegie, H. Ohadi, B. de Nijs, R. Esteban, J. Aizpurua, and J.J. Baumberg, "Single-molecule optomechanics in 'picocavities,'" *Science* 354 (2016): 726-729.

Berry, D., A.M. Childs, R. Cleve, R. Kothari, and R. Somma, "Simulating Hamiltonian dynamics with a truncated Taylor series," *Physical Review Letters* 114 (2015a): 090502.

Berry, D.W., A.M. Childs, and R. Kothari, "Hamiltonian Simulation with Nearly Optimal Dependence on all Parameters," *2015 IEEE 56th Annual Symposium on Foundations of Computer Science* (2015b).

Bittner, E.R., and C. Silva, "Estimating the conditions for polariton condensation in organic thin-film microcavities," *Journal of Chemical Physics* 136 (2012):034510.

Bloch, I., "Ultracold quantum gases in optical lattices," *Nature Physics* 1 (2005): 23-30.

Bloch, I., J. Dalibard, and S. Nascimbene, "Quantum simulations with ultracold quantum gases," *Nature Physics* 8 (2012): 267-276.

Boehme, C., and J. M. Lupton "Challenges for organic spintronics." *Nature Nanotechnology* 8 (2013): 612-5.

Boixo, S., S.V. Isakov, V.N. Smelyanskiy, R. Babbush, N. Ding, Z. Jiang, J.M. Martinis, and H. Neven, "Characterizing quantum supremacy in near-term devices." arXiv:1608.00263 (2016).

Brandao, F. and M. Kastoryano, "Finite correlation length implies efficient preparation of quantum thermal states." arXiv:1609.07877 (2016).

Bravyi, S., and A.Y. Kitaev, "Fermionic quantum computation," *Annals of Physics* 298 (2002): 210-226.

Bravyi, S., J.M. Gambetta, A. Mezzacapo, and K. Temme, "Tapering off qubits to simulate fermionic Hamiltonians," arXiv:1701.08213 (2017).

Britt, K.A., and T.S. Humble, "High-performance computing with quantum processing units." arXiv:1511.04386 (2015).

Brown, R. M., A. M. Tyryshkin, K. Porfyrakis, E. M. Gauger, B. W. Lovett, A. Ardavan, S. A. Lyon, G. A. D. Briggs, and J. J. L. Morton "Coherent State Transfer between an Electron and Nuclear Spin in $^{15}N@C_{60}$." *Physical Review Letters* 106 (2011): 110504.

Byrnes, T., N.Y. Kim, and Y. Yamamoto, "Exciton–polariton condensates," *Nature Physics* 10 (2014): 803-813.

Cao, Y., R. Babbush, J. Biamonte, and S. Kais, "Hamiltonian gadgets with reduced resource requirements," *Physical Review A* 91 (2015): 012315.

Carr, L.D., D. DeMille, R.V. Krems, and J. Ye, "Cold and ultracold molecules: science, technology and applications," *New Journal of Physics* 11 (2009): 055049.

Casey, S.R., and J.R. Sparks, "Vibrational Strong Coupling of Organometallic Complexes," *The Journal of Physical Chemistry C* 120 (2016): 28138-28143.

Chernick, E.T., Q. Mi, R.F. Kelley, E.A. Weiss, B.A. Jones, T.. Marks, M.A. Ratner, and M.R. Wasielewski, "Electron Donor-Bridge-Acceptor Molecules with Bridging Nitronyl Nitroxide Radicals: Influence of a Third Spin on Charge- and Spin-Transfer Dynamics," *Journal of the American Chemical Society* 128 (2006): 4356-4364.

Csehi, A., G.J. Halász, L.S. Cederbaum, and Á. Vibók, "Competition between Light-Induced and Intrinsic Nonadiabatic Phenomena in Diatomics," *The Journal of Physical Chemistry Letters* 8 (2017): 1624-1630.

Castelvecch, D., "IBM's quantum cloud computer goes commercial," *Nature* 543 (2017): 159.

Chang, Y-P., K. Długołęcki, J. Küpper, D. Rösch, D. Wild, and S. Willitsch, "Specific chemical reactivities of spatially separated 3-aminophenol conformers with cold Ca+ ions," *Science* 342 (2013): 98-101.

Chikkaraddy, R., B. de Nijs, F. Benz, S.J. Barrow, O.A. Scherman, E. Rosta, A. Demetriadou, P. Fox, O. Hess, and J.J. Baumberg, "Single-molecule strong coupling at room temperature in plasmonic nanocavities," *Nature* 535 (2016): 127–130.

Coles, D.M., N. Somaschi, P. Michetti, C. Clark, P.G. Lagoudakis, P.G. Savvidis, and D.G. Lidzey, "Polariton–mediated energy transfer between organic dyes in a strongly coupled optical microcavity," *Nature Materials* 13 (2014a): 712–719.

Coles, D.M., Y. Yang, Y. Wang, R.T. Grant, R.A. Taylor, S.K. Saikin, A. Aspuru-Guzik, D.G. Lidzey, J.K-H. Tang, and J.M. Smith, "Strong coupling between chlorosomes of photosynthetic bacteria and a confined optical cavity mode," *Nature Communications* 5 (2014b).

Colvin, M. T., R. Carmieli, T. Miura, S. Richert, D. M. Gardner, A. L. Smeigh, S. M. Dyar, S. M. Conron, M. A. Ratner, and M.R. Wasielewski, "Electron Spin Polarization Transfer from Photogenerated Spin-Correlated Radical Pairs to a Stable Radical Observer Spin," *The Journal of Physical Chemistry A* 117 (2013): 5314-5325.

Córcoles, A.D., E. Magesan, S.J. Srinivasan, A.W. Cross, M. Steffen, J.M. Gambetta, and J.M. Chow, "Demonstration of a quantum error detection code using a square lattice of four superconducting qubits," *Nature Communications* 6 (2015).

Cortés, E., W. Xie, J. Cambiasso, A.S. Jermyn, R. Sundararaman, P. Narang, S. Schlücker, and S.A. Maier, "Plasmonic hot electron transport drives nano-localized chemistry," *Nature Communications* 8 (2017).

D-Wave Systems Inc., "D-Wave 2000Q Technology Overview," (2017). URL: https://www.dwavesys.com/sites/default/files/D-Wave%202000Q%20Tech%20Collateral_0117F2.pdf

Dallaire-Demers, P–L., and F.K. Wilhelm, "Method to efficiently simulate the thermodynamic properties of the Fermi-Hubbard model on a quantum computer," *Physical Review* A 93 (2016a): 032303.

Dallaire-Demers, P–L., and F.K. Wilhelm, "Quantum gates and architecture for the quantum simulation of the Fermi-Hubbard model," *Physical Review* A 94 (2016b): 062304.

Dorfman, K.E., F. Schlawin, and S. Mukamel, "Stimulated Raman Spectroscopy with Entangled Light; Enhanced Resolution and Pathway Selection," *The Journal of Physical Chemistry Letters* 5 (2014): 2843-2849.

Dorfman, K.E., F. Schlawin, and S. Mukamel, "Nonlinear optical signals and spectroscopy with quantum light," *Reviews of Modern Physics* 88 (2016): 045008.

Du, J., N. Xu, X. Peng, P. Wang, S. Wu, and D. Lu. "NMR implementation of a molecular hydrogen quantum simulation with adiabatic state preparation." *Physical Review Letters* 104 (2010): 030502.

Duvenaud, D., D. Maclaurin, J. Aguilera-Iparraguirre, R. Gómez-Bombarelli, T. Hirzel, A. Aspuru-Guzik, and R.P. Adams, "Convolutional Networks on Graphs for Learning Molecular Fingerprints," *Conference on Neural Information Processing Systems* 28 (2015): 2215–2223.

Farhi, E., J. Goldstone and S. Gutmann, and M. Sipser, "Quantum Computation by Adiabatic Evolution," arXiv:quant-ph/0001106 (2000).

Farhi, E., J. Goldstone, and S. Gutmann, "A quantum approximate optimization algorithm," arXiv:1411.4028 (2014a).

Farhi, E., J. Goldstone, S. Gutmann, and H. Neven, "Quantum Algorithms for Fixed Qubit Architectures," arXiv:1703.06199 (2017).

Farhi, E., J. Goldstone, and S. Gutmann, "A quantum approximate optimization algorithm applied to a bounded occurrence constraint problem," arXiv:1412.6062 (2014b).

Farhi, E. and A.W. Harrow, "Quantum Supremacy through the Quantum Approximate Optimization Algorithm," arXiv:1602.07674 (2016).

Fataftah, M.S., J.M. Zadrozny, S.C. Coste, M.J. Graham, D.M. Rogers, and D.E. Freedman, "Employing Forbidden Transitions as Qubits in a Nuclear Spin-Free Chromium Complex," *Journal of the American Chemical Society* 138 (2016): 1344-1348.

Feynman, R.P., "Quantum mechanical computers," *Foundations of Physics* 16 (1986): 507-31.

Flick, J., M. Ruggenthaler, H. Appel, and A. Rubio, "Atoms and molecules in cavities, from weak to strong coupling in quantum-electrodynamics (QED) chemistry," *Proceedings of the National Academy of Sciences* 114 (2017): 3026-3034.

Forn-Díaz, P., J.J. García-Ripoll, B. Peropadre, J–L. Orgiazzi, M.A. Yurtalan, R. Belyansky, C.M. Wilson, and A. Lupascu, "Ultrastrong coupling of a single artificial atom to an electromagnetic continuum in the nonperturbative regime," *Nature Physics* 13 (2016): 39–43.

Frediani, L. and D. Sundholm, "Real-space numerical grid methods in quantum chemistry," *Physical Chemistry Chemical Physics* 17 (2015): 31357-31359.

Galbiati, M., L. Ferrier, D.D. Solnyshkov, D. Tanese, E. Wertz, A. Amo, M. Abbarchi, P. Senellart, I. Sagnes, A. Lemaître, E. Galopin, G. Malpuech, and J. Bloch, "Polariton Condensation in Photonic Molecules", *Physical Review Letters* 108 (2013): 126403.

Galego, J., F.J. Garcia-Vidal, and J. Feist, "Cavity-Induced Modifications of Molecular Structure in the Strong-Coupling Regime," *Physical Review X* 5 (2015): 041022.

Gallagher, N., A. Olankitwanit, and A. Rajca, "High-Spin Organic Molecules," *The Journal of Organic Chemistry* 80 (2015): 1291–1298.

Gharibian, S., Y. Huang, Z. Landau, and S.W. Shin, "Quantum Hamiltonian Complexity," *Foundations and Trends in Theoretical Computer Science* 10 (2015): 159-282.

Giovannetti, V., S. Lloyd, and L. Maccone, "Quantum random access memory," *Physical Review Letters* 100 (2008): 160501.

Goldman, N., J.C. Budich, and P. Zoller, "Topological quantum matter with ultracold gases in optical lattices," *Nature Physics* 12 (2016): 639–645.

Gómez-Bombarelli, R., D. Duvenaud, J.M. Hernández-Lobato, J. Aguilera-Iparraguirre, T.D. Hirzel, R.P. Adams, and A. Aspuru-Guzik, "Automatic chemical design using a data-driven continuous representation of molecules," arXiv:1610.02415 (2016).

Green, A.S., P.L. Lumsdaine, N.J. Ross, P. Selinger, and B. Valiron, "An Introduction to Quantum Programming in Quipper," arXiv:1304.5485 (2013).

Grosso, M., A. Lienhard, F. Efetov, F. Jarillo-Herrero, and E. Aharonovich, "Tunable and high purity room-temperature single photon emission from atomic defects in hexagonal boron nitride," arXiv:1611.03515 (2016).

Grover, L.K., "A fast quantum mechanical algorithm for database search," *Proceedings of the twenty-eighth annual ACM symposium on Theory of computing - STOC* 96 (1996).

Guerreschi, G.G., and M. Smelyanskiy, "Practical optimization for hybrid quantum-classical algorithms," arXiv:1701.01450 (2017).

Harrow, A.W., A. Hassidim, and S. Lloyd, "Quantum algorithm for solving linear systems of equations," *Physical Review Letters* 103 (2009): 150502.

Hasharoni, K., H. Levanon, S. R. Greenfield, D. J. Gosztola, W. A. Svec, and M. R. Wasielewski, "Mimicry of the Radical Pair and Triplet-States in Photosynthetic Reaction Centers with a Synthetic Model," *Journal of the American Chemical Society* 117 (1995): 8055-8056.

Hastings, M. B., D. Wecker, B. Bauer, and M. Troyer, "Improving Quantum Algorithms for Quantum Chemistry," arXiv:1403.1539 (2015).

Herrera, F., and F.C. Spano, "Cavity-Controlled Chemistry in Molecular Ensembles," *Physical Review Letters* 116 (2016): 238301.

Hinton, G., and R. Salakhutdinov, "Reducing the Dimensionality of Data with Neural Networks," *Science* 28 (2006): 504-507.

Höfer, M., L. Riegger, F. Scazza, C. Hofrichter, D.R. Fernandes, M.M. Parish, J. Levinsen, I. Bloch, and S. Fölling, "Observation of an Orbital Interaction-Induced Feshbach Resonance in Yb 173," *Physical Review Letters* 115 (2015): 265302.

Hoff, A.J., P. Gast, S.A. Dzuba, C.R. Timmel, C.E. Fursman, and P.J. Hore, "The nuts and bolts of distance determination and zero- and double-quantum coherence in photoinduced radical pairs," *Spectrochimica Acta* 54A (1998): 2283-2293.

Hosten, O., N.J. Engelsen, R. Krishnakumar, and M.A. Kasevich, "Measurement noise 100 times lower than the quantum-projection limit using entangled atoms," *Nature* 529 (2016): 505–508.

Huh, J., G.G. Guerreschi, B. Peropadre, J.R. McClean, and A. Aspuru-Guzik, "Boson sampling for molecular vibronic spectra," *Nature Photonics* 9 (2015): 615–620.

Humble, T.S., A.J. McCaskey, J. Schrock, H. Seddiqi, K.A. Britt, and N. Imam, "Performance Models for Split-execution Computing Systems," *Parallel and Distributed Processing Symposium Workshops* (2016): 545-554.

Jain, H., and P. Ong, "New opportunities for materials informatics: Resources and data mining techniques for uncovering hidden relationships," *Journal of Materials Research* 31 (2016): 977-994.

Jones, N. C., Whitfield, J. Dl, McMachon, P. L., Yung, M.-H., Van Meter, R., Aspuru-Guzik, A. and Y. Yamamoto, "Faster quantum chemistry simulation on fault-tolerant quantum computers" *The New Journal of Physics* 14, (2012): 115023.

Juffmann, T., S.A. Koppell, B.B. Klopfer, C. Ophus, R. Glaeser, and M.A. Kasevich, "Multi-pass transmission electron microscopy," arXiv:1612.04931 (2016).

Kais, S., A.R. Dinner, and S.A. Rice, eds, "*Quantum Information and Computation for Chemistry: Advances in Chemical Physics Vol. 154*," Hoboken, NJ: J. Wiley & Sons (2014).

Kajita, M., G.Gopakumar, M. Abe, M. Hada, and M. Keller, "Test of m p/m e changes using vibrational transitions in N 2+," *Physical Review A* 89 (2014): 032509.

Kandala, A., A. Mezzacapo, K. Temme, M. Takita, J.M. Chow, and J.M. Gambetta, "Hardware-efficient Quantum Optimizer for Small Molecules and Quantum Magnets," arXiv:1704.05018 (2017).

Kandrashkin, Y. E., and K. M. Salikhov "Numerical Simulation of Quantum Teleportation Across Biological Membrane in Photosynthetic Reaction Centers." *Applied Magnetoc Resonance* 37 (2010): 549-566.

Kassal, I., S.P. Jordan, P.J. Love, M. Mohseni, and A. Aspuru-Guzik, "Polynomial-time Quantum Algorithm for the Simulation of Chemical Dynamics," *Proceedings of the National Academy of Sciences* 105 (2008): 18681-18686.

Kasprzak, J., M. Richard, S. Kundermann, A. Baas, P. Jeambrun, J.M. J. Keeling, F.M. Marchetti, M.H. Szymańska, R. André, J.L. Staehli, V. Savona, P.B. Littlewood, B. Deveaud, and L.S. Dang, "Bose–Einstein condensation of exciton polaritons," *Nature* 443 (2006): 409.

Kaufman, A. M., B.J. Lester, M. Foss-Feig, M.L. Wall, A.M. Rey, and C.A. Regal, "Entangling two transportable neutral atoms via local spin exchange," *Nature* 527 (2015): 208-211.

Kelly, J., R. Barends, A.G. Fowler, A. Megrant, E. Jeffrey, T.C. White, D. Sank et al., "State preservation by repetitive error detection in a superconducting quantum circuit," *Nature* 519 (2015): 66-69.

Kéna-Cohen, S. and S.R. Forrest, "Room-temperature polariton lasing in an organic single-crystal microcavity," *Nature Photonics* 4 (2010): 371-375.

Khanyile, N.B., G. Shu, and K.R. Brown, "Observation of vibrational overtones by single-molecule resonant photo-dissociation," *Nature Communications* 6 (2015).

Kieferova, M., and N. Wiebe, "Tomography and Generative Data Modeling via Quantum Boltzmann Training," arXiv:1612.05204 (2016).

Kirk, M.L., D.A. Shultz, R.D. Schmidt, D. Habel-Rodriguez, H. Lee, and J. Lee, "Ferromagnetic Nanoscale Electron Correlation Promoted by Organic Spin-Dependent Delocalization," *Journal of the American Chemical Society* 131 (2009): 18304-18313.

Kitaev, A., "Anyons in an exactly solved model and beyond," *Annals of Physics* 321 (2006): 2-111.

Kivlichan, I.D., N. Wiebe, R. Babbush, A. Aspuru-Guzik, "Bounding the costs of quantum simulation of many-body physics in real space," arXiv:1608.05696 (2016).

Kobr, L., D.M. Gardner, A.L. Smeigh, S.M. Dyar, S.D. Karlen, R. Carmieli, and M.R. Wasielewski, "Fast Photodriven Electron Spin Coherence Transfer: A Quantum Gate Based on a Spin Exchange J-Jump," *Journal of the American Chemical Society* 134 (2012): 12430-12433.

Kothe, G., S. Weber, E. Ohmes, M. C. Thurnauer, and J. R. Norris "High Time Resolution Electron-Paramagnetic-Resonance of Light-Induced Radical Pairs in Photosynthetic Bacterial Reaction Centers - Observation of Quantum Beats." *Journal of the American Chemical Society* 116 (1994): 7729-7734.

Kothe, G., T. Yago, J-U. Weidner, G. Link, M. Lukaschek, and T-S. Lin, "Quantum Oscillations and Polarization of Nuclear Spins in Photoexcited Triplet States," *The Journal of Physical Chemistry B* 114 (2010): 14755-14762.

Kohstall, C., M. Zaccanti, M. Jag, A. Trenkwalder, P. Massignan, G.M. Bruun, F. Schreck, and R. Grimm, "Metastability and coherence of repulsive polarons in a strongly interacting Fermi mixture," *Nature* 485 (2012): 615–618.

Kowalewski, M., K. Bennett, and S. Mukamel, "Non-adiabatic dynamics of molecules in optical cavities," *The Journal of Chemical Physics* 144 (2016a): 054309.

Kowalewski, M., K. Bennett, and S. Mukamel, "Cavity Femtochemistry: Manipulating Nonadiabatic Dynamics at Avoided Crossings," *The Journal of Physical Chemistry Letters* 7 (2016b): 2050-2054.

Krzyaniak, M.D., L. Kobr, B.K. Rugg, B.T. Phelan, E.A. Margulies, J.N. Nelson, R.M. Young, and M.R. Wasielewski "Fast photo-driven electron spin coherence transfer: the effect of electron-nuclear hyperfine coupling on coherence dephasing." *Journal of Materials Chemistry C* 3 (2015): 7962-7967.

Landau, L.D.,"Theory of the Superfluidity of Helium II," *Physical Review* 60 (1941): 356-358.

Lanyon, B.P., J.D. Whitfield, G.G. Gillett, M.E. Goggin, M.P. Almeida, I. Kassal, J.D. Biamonte et al. "Towards quantum chemistry on a quantum computer." *Nature Chemistry* 2, no. 2 (2010): 106-111.

Laukenmann, K., S. Weber, G. Kothe, C. Oesterle, A. Angerhofer, M.R. Wasielewski, W.A. Svec, and J.R. Norris "Quantum Beats of the Radical Pair State in Photosynthetic Models Observed by Transient Electron Paramagnetic Resonance." *The Journal of Physical Chemistry* 99 (1995): 4324-4329.

Leuenberger, M.N., and D. Loss, "Quantum computing in molecular magnets," *Nature* 410 (2001): 789-793.

Leyton-Brown, K., H.H. Hoos, F. Hutter, and L. Xu, "Understanding the empirical hardness of NP-complete problems," *Communications of the ACM* 57 (2014): 98-107.

Li, X., and W.V. Liu, "Physics of higher orbital bands in optical lattices: a review," *Reports on Progress in Physics* 79 (2016): 116401.

Lin, C.Y.Y., and Y. Zhu, "Performance of QAOA on Typical Instances of Constraint Satisfaction Problems with Bounded Degree," arXiv:1601.01744 (2016).

Lin, T-S., D.J. Sloop, and C-Y. Mou, "Utilization of polarized electron spin of organic molecules in quantum computing," Quantum Information Science, *Proceedings of the Asia-Pacific Conference, 1st, Tainan, Taiwan, Dec. 10-13, 2004* (2005): 205-213.

Lloyd, S., "Universal Quantum Simulators," *Science* 273 (1996): 1073-1078.

Lloyd, S., S. Garnerone, and P. Zanardi, "Quantum algorithms for topological and geometric analysis of data," *Nature Communications* 7 (2016): 10138.

Lühmann, D-S., C.Weitenberg, and K. Sengstock, "Emulating Molecular Orbitals and Electronic Dynamics with Ultracold Atoms," *Physical Review X* 5 (2015): 031016.

Martínez-Martínez, L., R.F. Ribeiro, J.A. Campos-González-Angulo, and J. Yuen-Zhou, "Can ultrastrong coupling change ground-state chemical reactions?", (to be submitted 2017).

Matsugi, A., H. Shiina, T. Oguchi, and K. Takahashi, "Time-Resolved Broadband Cavity-Enhanced Absorption Spectroscopy behind Shock Waves," *The Journal of Physical Chemistry A* 120 (2016): 2070-2077.

McClean, J.R., R. Babbush, P.J. Love, and A. Aspuru-Guzik, "Exploiting locality in quantum computation for quantum chemistry," *The Journal of Physical Chemistry Letters* 5 (2014): 4368-4380.

McClean, J.R., M.E. Kimchi-Schwartz, J. Carter, and W.A. de Jong, "Hybrid quantum-classical hierarchy for mitigation of decoherence and determination of excited states," *Physical Review A* 95 (2017): 042308.

McClean, J.R., J. Romero, R. Babbush, and A. Aspuru-Guzik, "The theory of variational hybrid quantum-classical algorithms," *New Journal of Physics* 18 (2016): 023023.

Mehring, M., J. Mende, and W. Scherer, "Entanglement between an Electron and a Nuclear Spin ½," *Physical Review Letters* 90 (2003): 153001.

Mertens, J., M-E. Kleemann, R. Chikkaraddy, P. Narang, and J.J. Baumberg, "How Light Is Emitted by Plasmonic Metals," *Nano Letters* 17 (2017): 2568–2574.

Mi, Q., E.T. Chernick, D.W. McCamant, E.A. Weiss, M.A. Ratner, and M.R. Wasielewski, "Spin Dynamics of Photogenerated Triradicals in Fixed Distance Electron Donor-Chromophore-Acceptor-TEMPO Molecules," *The Journal of Physical Chemistry A* 110 (2006): 7323-7333.

Miura, T., and M.R. Wasielewski, "Manipulating photogenerated radical ion pair lifetimes in wire-like molecules using microwave pulses: Molecular spintronic gates," *Journal of the American Chemical Society* 133 (2011): 2844-2847.

Mohseni, M., P. Read, H. Neve, S. Boixio, V. Denchev, R. Babbush, A. Fowler, V. Smelyanskiy, and J. Martinis, "Commercialize quantum technologies in five years," *Nature* 543 (2017): 171-175.

Moore, M.G. and A. Vardi, "Bose-Enhanced Chemistry: Amplification of Selectivity in the Dissociation of Molecular Bose-Einstein Condensates," *Physical Review Letters* 88 (2012): 160402.

Morton, J.J.L., A.M. Tyryshkin, R.M. Brown, S. Shankar, B.W. Lovett, A. Ardavan, T. Schenkel, E.E. Haller, J.W. Ager, and S.A. Lyon, "Solid-state quantum memory using the 31P nuclear spin," *Nature* 455 (2008): 1085-1088.

Motes, K.R., J.P. Olson, E.J. Rabeaux, J.P. Dowling, S.J. Olson, and P.P. Rohde, "Linear Optical Quantum Metrology with Single Photons: Exploiting Spontaneously Generated Entanglement to Beat the Shot-Noise Limit," *Physical Review Letters* 114 (2015): 170802.

Muallem, M., A. Palatnik, G.D. Nessim, and Y.R. Tischler, "Strong light-matter coupling and hybridization of molecular vibrations in a low-loss infrared microcavity," *The Journal of Physical Chemistry Letters* 7 (2016): 2002-2008.

Narang, P., R. Sundararaman, H.A. Atwater, "Plasmonic hot carrier dynamics in solid-state and chemical systems for energy conversion," *Nanophotonics* 5 (2016).

O'Malley, P.J.J., Ryan Babbush, I.D. Kivlichan, J. Romero, J.R. McClean, R. Barends, J. Kelly et al., "Scalable quantum simulation of molecular energies," *Physical Review X* 6 (2016): 031007.

Ochsenbein, S.T. and D.R. Gamelin, "Quantum oscillations in magnetically doped colloidal nanocrystals," *Nature Nanotechnology* 6 (2011): 112–115.

Olson, J.P., K.R. Motes, P.M. Birchall, N.M. Studer, M. LaBorde, T. Moulder, P.P. Rohde, and J.P. Dowling, "Linear optical quantum metrology with single photons — Experimental errors, resource counting, and quantum Cramér-Rao bounds," arXiv:1610.07128 (2016).

Ortiz, G., J.E. Gubernatis, E. Knill, and R. Laflamme, "Quantum algorithms for fermionic simulations," *Physical Review A* 64 (2001): 022319.

Oyaizu, K. and H. Nishide, "Radical Polymers for Organic Electronic Devices: A Radical Departure from Conjugated Polymers?" *Advanced Materials* 21 (2009): 2339-2344.

Pachón, L.A., Marcus, A.H., and A. Aspuru-Guzik, "Quantum Process Tomography by 2D Fluorescence Spectroscopy," *The Journal of Chemical Physics* 142 (2015): 212442.

Perdomo-Ortiz, A., N. Dickson, M. Drew-Brook, G. Rose, and A. Aspuru-Guzik, "Finding low-energy conformations of lattice protein models by quantum annealing," *Scientific Reports* 2 (2012).

Peruzzo, A., J.R. McClean, P. Shadbolt, M-H. Yung, X-Q. Zhou, P.J. Love, A. Aspuru-Guzik, and J.L. O'brien, "A variational eigenvalue solver on a quantum processor," arXiv:1304.3061 (2013).

Plumhof, J.D., T. Stöferle, L. Mai, U. Scherf, and R.F. Mahrt, "Room-temperature Bose–Einstein condensation of cavity exciton–polaritons in a polymer," *Nature Materials* 13 (2014): 247-252.

Poulin, D., M.B. Hastings, D. Wecker, N. Wiebe, A.C. Doherty, and M. Troyer, "The Trotter Step Size Required for Accurate Quantum Simulation of Quantum Chemistry," arXiv:1406.4920 (2015).

Racah, E., S. Ko, P. Sadowski, W. Bhimji, C. Tull, S-Y. Oh, P. Baldi, and Prabhat, "Revealing Fundamental Physics from the Daya Bay Neutrino Experiment Using Deep Neural Networks," *2016 15th IEEE International Conference on Machine Learning and Applications* (2016).

Raymer, M.R., A.H. Marcus, J.R. Widom, and D.L.P. Vitullo, "Entangled Photon-Pair Two Dimensional Fluorescence Spectroscopy (EPP-2DFS)," *Journal of Physical Chemistry B* 117 (2013): 15559-15575.

Rebentrost, P., M. Mohseni, and S. Lloyd, "Quantum support vector machine for big data classification," *Physical Review Letters* 113 (2014): 130503.

Reiher, M., N. Wiebe, K.M. Svore, D. Wecker, and M. Troyer, "Elucidating reaction mechanisms on quantum computers," arXiv:1605.03590 (2016).

Renner, J., A. Farbin, J.M. Vidal, J.M. Benlloch-Rodríguez, A. Botas, P. Ferrario, J.J. Gómez-Cadenas, V. Álvarez et al., "Background rejection in NEXT using deep neural networks," arXiV:1609.06202 (2016).

Ribeiro, R. and J. Yuen-Zhou, "Continuous vibronic symmetries in Jahn-Teller models," arXiV: 1705.08104 (2017).

Romero, J., R. Babbush, J.R. McClean, C. Hempel, P. Love, and A. Aspuru-Guzik, "Strategies for quantum computing molecular energies using the unitary coupled cluster ansatz," arXiv:1701.02691 (2017).

Romero, J., J. Olson, and A. Aspuru-Guzik, "Quantum autoencoders for efficient compression of quantum data," arXiv:1612.02806 (2016).

Rubin, N.C., "A Hybrid Classical/Quantum Approach for Large-Scale Studies of Quantum Systems with Density Matrix Embedding Theory," arXiv:1610.06910 (2016).

Rugar, D., "Single spin detection by magnetic resonance force microscopy," *Nature* 430 (2004): 329–332.

Saha, S., A. Rakshit, D. Chakraborty, A. Pal, and B. Deb, "Optical Feshbach resonances through a molecular dark state: Efficient manipulation of p-wave resonances in fermionic Yb 171 atoms," *Physical Review A* 90 (2014): 012701.

Salikhov, K.M., J.H. Golbeck, and D. Stehlik, "Quantum teleportation across a biological membrane by means of correlated spin pair dynamics in photosynthetic reaction centers," *Applied Magnetic Resonance* 31 (2007): 237–252.

Santagati, R., J. Wang, A.A. Gentile, S. Paesani, N. Wiebe, J.R. McClean, S.R. Short et al., "Quantum simulation of Hamiltonian spectra on a silicon chip," arXiv:1611.03511 (2016).

Sato, K., S. Nakazawa, R. Rahimi, T. Ise, S. Nishida, T. Yoshino, N. Mori, K. Toyota, D. Shiomi, Y. Yakiyama, Y. Morita, M. Kitagawa, K. Nakasuji, M. Nakahara, H. Hara, P. Carl, P. Hoefer, and T. Takui, "Molecular electron-spin quantum computers and quantum information processing: pulse-based electron magnetic resonance spin technology applied to matter spin-qubits," *Journal of Materials Chemistry* 19 (2009): 3739–3754.

Sawaya, N., M. Smelyanskiy, J.R. McClean, and A. Aspuru-Guzik, "Error sensitivity to environmental noise in quantum circuits for chemical state preparation," *Journal of Chemical Theory and Computation* 12 (2016): 3097–3108.

Schirhagl, R., K. Chang, M. Loretz, and C.L. Degen, "Nitrogen-vacancy centers in diamond: nanoscale sensors for physics and biology," *Annual Review of Physical Chemistry* 65 (2014): 83–105.

Schlawin, F., K.E. Dorfman, B.P. Fingerhut, and S. Mukamel, "Suppression of Population transport and Control of Exciton Distributions by Entangled Photons," *Nature Communications* 4 (2013): 1782.

Schmidt, P.O., T. Rosenband, C. Langer, W.M. Itano, J.C. Bergquist, and D.J. Wineland, "Spectroscopy using quantum logic," *Science* 309 (2005): 749–752.

Schröder, T., M.E. Trusheim, M. Walsh, L. Li, J. Zheng et al., "Scalable Focused Ion Beam Creation of Nearly Lifetime-Limited Single Quantum Emitters in Diamond Nanostructures," arXiv:1610.09492 (2016).

Schuch, N. and F. Verstraete, "Computational complexity of interacting electrons and fundamental limitations of density functional theory," *Nature Physics* 5 (2009): 732–735.

Schuld, M., I. Sinayskiy, and F. Petruccione, "Prediction by linear regression on a quantum computer," *Physical Review A* 94 (2016): 022342.

Schweiger, A. and G. Jeschke, *Principles of pulse electron paramagnetic resonance*. 1 ed. (Oxford, Oxford University Press, 2001).

Shabani, A., J. Roden, and K.B. Whaley, "Continuous Measurement of a Non-Markovian Open Quantum System," *Physical Review Letters* 112 (2014): 113601.

Shen, Y., X. Zhang, S. Zhang, J-N. Zhang, M-H. Yung, and K. Kim, "Quantum implementation of the unitary coupled cluster for simulating molecular electronic structure," *Physical Review A* 95 (2017): 020501.

Shiddiq, M., D. Komijani, Y. Duan, A. Gaita-Ariño, E. Coronado, and S. Hill, "Enhancing coherence in molecular spin qubits via atomic clock transitions," *Nature* 531 (2016): 348–351.

Shor, P.W., "Polynomial-Time Algorithms for Prime Factorization and Discrete Logarithms on a Quantum Computer," *SIAM Review* 41 (1999): 303-32.

Simmons, S., R.M. Brown, H.Riemann, N.V. Abrosimov, P. Becker, H-J. Pohl, M.L.W. Thewalt, K.M. Itoh, and J.J.L.

Morton, "Entanglement in a solid-state spin ensemble," *Nature* 470 (2011): 69-72.

Smelyanskiy, M., N. Sawaya, and A. Aspuru-Guzik, "qHiPSTER: The Quantum High Performance Software Testing Environment," arXiv:1601.07195 (2016).

Smith, R.S., M.J. Curtis, and W.J. Zeng, "A Practical Quantum Instruction Set Architecture," arXiv:1608.03355 (2016).

Steiger, D.S., T. Häner, and M. Troyer, "ProjectQ: An Open Source Software Framework for Quantum Computing," arXiv:1612.08091 (2016).

Sun, D., E. Ehrenfreund, and Z.V. Vardeny, "The first decade of organic spintronics research," *Chemical Communications* 50 (2014): 1781-93.

Tarruell, L., D. Greif, T. Uehlinger, G. Jotzu, and T. Esslinger, "Creating, moving and merging Dirac points with a Fermi gas in a tunable honeycomb lattice," *Nature* 483 (2012): 302–305.

Tekavec, P.F., G.A. Lott, and A.H. Marcus, "Fluorescence-Detected Two-Dimensional Electronic Coherence Spectroscopy by Acousto-Optic Phase Modulation," *The Journal of Chemical Physics* 127 (2007): 214307.

Tran, T.T., K. Bray, M. Ford, M. Toth, and I. Aharonovich, "Quantum emission from hexagonal boron nitride monolayers," *Nature Nanotechnology* 11 (2016): 37–41.

Troiani, F., A. Ghirri, M. Affronte, S. Carretta, P. Santini, G. Amoretti, S. Piligkos, G. Timco, and R.E.P. Winpenny, "Molecular Engineering of Antiferromagnetic Rings for Quantum Computation," *Physical Review Letters* 94 (2005): 207208.

Vasa, P., W. Wang, R. Pomraenke, M. Lammers, M. Maiuri, C. Manzoni, G. Cerullo, and C. Lienau, "Real-time observation of ultrafast Rabi oscillations between excitons and plasmons in metal nanostructures with J-aggregates," *Nature Photonics* 7 (2013): 128–132.

Veis, L., J. Višňák, T. Fleig, S. Knecht, T. Saue, L. Visscher, and J. Pittner, "Relativistic quantum chemistry on quantum computers," *Physical Review A* 85 (2012): 030304(R).

Wang, H., Y. He, Y-H. Li, Z-E. Su, B. Li, H–L. Huang, X. Ding, M-C. Chen, C. Liu, J. Qin, J-P. Li, Y-M. He, C. Schneider, M. Kamp, C-Z. Peng, S. Höfling, C-Y. Lu, and J-W. Pan, "High-efficiency multiphoton boson sampling," *Nature Photonics* (2017).

Wang, H., S. Kais, A. Aspuru-Guzik, and M.R. Hoffmann, "Quantum algorithm for obtaining the energy spectrum of molecular systems," *Physical Chemistry Chemical Physics* 10 (2008): 5388.

Wang, Y., F. Dolde, J. Biamonte, R. Babbush, V. Bergholm, S. Yang, I. Jakobi et al., "Quantum simulation of helium hydride cation in a solid-state spin register," *ACS Nano* 9 (2015): 7769-7774.

Wecker, D., B. Bauer, B.K. Clark, M.B. Hastings, and M. Troyer, "Gate-count estimates for performing quantum chemistry on small quantum computers," *Physical Review A* 90 (2014a): 022305.

Wecker, D., M.B. Hastings, and M. Troyer, "Progress towards practical quantum variational algorithms," *Physical Review A* 92 (2015): 042303.

Wecker, D., M.B. Hastings, and M. Troyer, "Training a quantum optimizer," *Physical Review A* 94 (2016): 022309.

Wecker, D. and K. Svore, "LIQUi|>: A Software Design Architecture and Domain-Specific Language for Quantum Computing," arXiv:1402.4467 (2014b).

Welch, J., D. Greenbaum, S. Mostame, and A. Aspuru-Guzik, "Efficient quantum circuits for diagonal unitaries without ancillas," *New Journal of Physics* 16 (2014): 033040.

Wesenberg, J.H., A. Ardavan, G.A. Briggs, J.J. Morton, R.J. Schoelkopf, D.I. Schuster, and K. Molmer, "Quantum computing with an electron spin ensemble," *Physical Review Letters* 103 (2009): 070502.

Whitfield, J.D., J. Biamonte, and A. Aspuru-Guzik, "Simulation of Electronic Structure Hamiltonians Using Quantum Computers," *Molecular Physics* 109 (2011): 735–750.

Whitfield, J.D., P.J. Love, and A. Aspuru-Guzik, "Computational Complexity in Electronic Structure," *Physical Chemistry Chemical Physics* 15 (2013): 397-411.

Wiebe, N., D. Braun, and S. Lloyd, "Quantum data fitting," *Physical Review Letters* 109 (2012): 050505.

Wiesner, S., "Simulations of Many-body Quantum Systems by a Quantum Computer," arXiv:quant-ph/9603028 (1996).

Wolf, F., Y. Wan, J.C. Heip, F. Gebert, C. Shi, and P.O. Schmidt, "Non-destructive state detection for quantum logic spectroscopy of molecular ions," *Nature* 530 (2016): 457–460.

Wolfowicz, G., A. Tyryshkin, R.E. George, H. Riemann, N.V. Abrosimov, P. Becker, H-J. Pohl, M. Thewalt, S.A. Lyon, and J.J.L. Morton, "Atomic clock transitions in silicon-based spin qubits," *Nature Nanotechnology* 8 (2013): 561–564.

Yamamoto, S., S. Nakazawa, K. Sugisaki, K. Sato, K. Toyota, D. Shiomi, and T. Takui, "Adiabatic quantum computing with spin qubits hosted by molecules," *Physical Chemistry Chemical Physics* 17 (2015): 2742–2749.

Yang, Z-C., A. Rahmani, A. Shabani, H. Neven, and C. Chamon, "Optimizing Variational Quantum Algorithms using Pontryagin's Minimum Principle," arXiv:1607.06473 (2016).

Yu, C., M.J. Graham, J.M. Zadrozny, J. Niklas, M. Krzyaniak, M.R. Wasielewski, O.G. Poluektov, and D.E. Freedman, "Long Coherence Times in Nuclear Spin-Free Vanadyl Qubits," *Journal of the American Chemical Society* 138 (2016): 14678–14685.

Yuen-Zhou, J., D.H. Arias, D.M. Eisele, C.P. Steiner, J.J. Krich, M. Bawendi, K.A. Nelson, and A. Aspuru-Guzik, "Coherent exciton dynamics in supramolecular light-harvesting nanotubes revealed by ultrafast quantum process tomography." *ACS Nano* 8 (2014a): 5527-5534.

Yuen-Zhou, J., J.J. Krich, I. Kassal, A.S. Johnson, and A. Aspuru-Guzik, "Ultrafast Spectroscopy: Quantum information and wavepackets," *IOP Publishing* (2014b).

Yuen-Zhou, J., J.J. Krich, M. Mohseni, and A. Aspuru-Guzik, "Quantum State and Process Tomography of Energy Transfer Systems via Ultrafast Spectroscopy," *Proceedings of the National Academy of Sciences* 108 (2011): 17615–17620.

Yuen-Zhou, J., S.K. Saikin, N.Y. Yao, and A. Aspuru-Guzik, "Topologically protected excitons in porphyrin thin films," *Nature Materials* 13 (2014c): 1026-1032.

Yuen-Zhou, J., S.K. Saikin, T. Zhu, M. Onbalsi, C. Ross, V. Bulovic, and M. Baldo, "Plexcitons: Dirac points and topological modes," *Nature Communications* 7 (2016): 11783.

Yung, M-H., J. Casanova, A. Mezzacapo, J. McClean, L. Lamata, A. Aspuru-Guzik, and E. Solano, "From transistor to trapped-ion computers for quantum chemistry," *Scientific Reports* 4 (2014).

Zadrozny, J.M., J. Niklas, O.G. Poluektov, and D.E. Freedman, "Millisecond Coherence Time in a Tunable Molecular Electronic Spin Qubit," *ACS Central Science* 1 (2015): 488–492.

Zalka, Christof, "Efficient simulation of quantum systems by quantum computers," *Proceedings of the Royal Society of London A* 454 (1998): 313-322.

Zhang, Y., T. Grover, A. Turner, M. Oshikawa, and A. Vishwanath, "Quasiparticle statistics and braiding from ground-state entanglement," *Physical Review B* 85 (2012): 235151.

Zhao, Z., J.K. Fitzsimons, and J.F. Fitzsimons, "Quantum assisted Gaussian process regression," arXiv:1512.03929 (2015).

Zhou, S., M. Yamamoto, G.A.D. Briggs, H. Imahori, and K. Porfyrakis, "Probing the Dipolar Coupling in a Heterospin Endohedral Fullerene–Phthalocyanine Dyad," *Journal of the American Chemical Society* 138 (2016): 1313–1319.

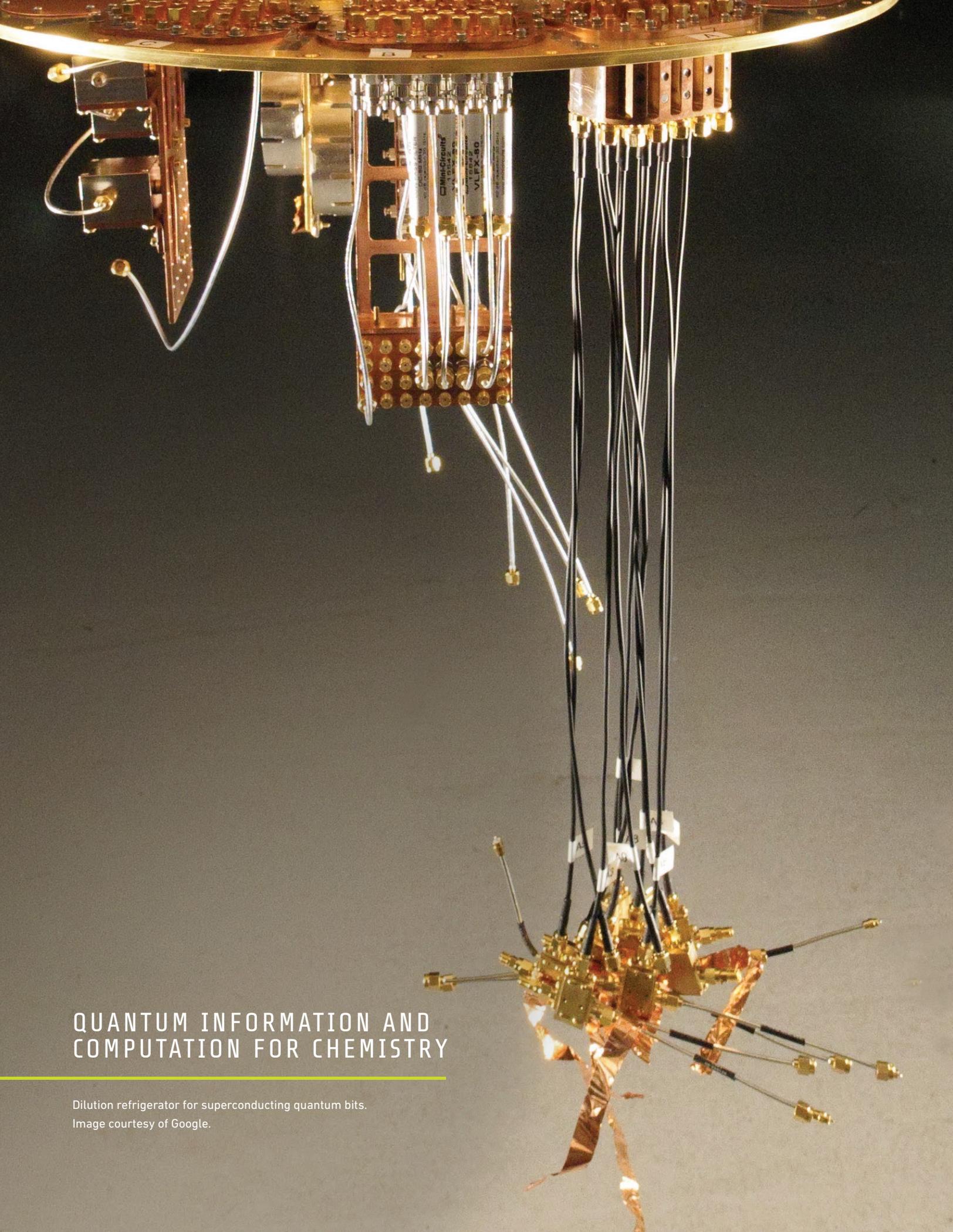

QUANTUM INFORMATION AND
COMPUTATION FOR CHEMISTRY

Dilution refrigerator for superconducting quantum bits.
Image courtesy of Google.